\documentclass[]{article}
\usepackage{amsmath,mathtools,amssymb,mathrsfs}
\usepackage[utf8]{inputenc}
\usepackage[dvipsnames,table,xcdraw]{xcolor}
\usepackage{todonotes}
\usepackage{geometry}
	
\usepackage{longtable}
\usepackage{titling}
\usepackage{array}
\usepackage{ragged2e}

\usepackage[title]{appendix}
\usepackage{authblk}
\usepackage{caption}
\usepackage[shortcuts]{extdash}
\usepackage{subcaption}
\usepackage{titlesec}
\titlespacing*{\section}
{0pt}{1.5ex plus .8ex minus .2ex}{0.7ex plus .2ex}
\titlespacing*{\subsection}
{0pt}{1ex plus .5ex minus .2ex}{0.4ex plus .1ex}
\titlespacing*{\subsubsection}
{0pt}{.8ex plus .4ex minus .2ex}{0.2ex plus .1ex}

\usepackage{tikz}
\usetikzlibrary{arrows,decorations.markings,plotmarks}
\usetikzlibrary{arrows.meta}

\usepackage{wrapfig}
\usepackage{multirow}
\usepackage[numbers,sort&compress]{natbib}

\usepackage{parskip} 

\graphicspath{{./Files/Figures}}

\usepackage{hyperref}
\hypersetup{
	linktocpage,
	colorlinks,
	urlcolor=red,	
	citecolor=ForestGreen,
	linkcolor=blue,
	filecolor=blue,
	pdfnewwindow,
	pdftitle={What is a DOF in classical mechanics?},
	pdfauthor={DLH,IB,JMB},
	pdfsubject={DOFs}}

\newcommand{\R}{\mathbb{R}}
\renewcommand{\S}{\mathbb{S}}

\newtheorem{definition}{Definition}

\DeclareRobustCommand{\rchi}{{\mathpalette\irchi\relax}}
\newcommand{\irchi}[2]{\raisebox{\depth}{$#1\chi$}} 
            
\DeclareRobustCommand{\straight}{\!\!
\raisebox{-0.25ex}{
\begin{tikzpicture}[
    middle/.style={
        decoration={
            markings,
            mark=at position 0.4 with{\pgfuseplotmark{|}},
        },
        postaction=decorate,
    }
]
\draw[middle,-{Latex[length=.7mm, width=1mm]}] (0,0)--(.15,0.25);
\end{tikzpicture}}}
     
\DeclareRobustCommand{\antistraight}{\!\!
\scalebox{1.2}{\raisebox{-0.25ex}{
\begin{tikzpicture}[
    end/.style={
        decoration={
            markings,
            mark=at position 1 with{\pgfuseplotmark{|}},
        },
        postaction=decorate,
    }
]
\draw[end,double,{Latex[length=.7mm, width=1mm]}-] (0,0)--(.1,0.15);
\end{tikzpicture}}}}

\DeclareRobustCommand{\clock}{\!\!
\raisebox{0.25ex}{
\begin{tikzpicture}
\draw (0,0)--(.155,0);
\draw[-{Latex[length=.7mm, width=1mm]}] (.155,0)--(.25,0.145);
\end{tikzpicture}}}

\DeclareRobustCommand{\counterclock}{\!\!
\raisebox{0.1ex}{
\begin{tikzpicture}
\draw (0,0)--(.075,0.125);
\draw[-{Latex[length=.7mm, width=1mm]}] (.075,0.125)--(.25,0.125);
\end{tikzpicture}}}

\title{What is a degree of freedom?\\[-.5ex] Configuration spaces and their topology}
\author{} 

 \date{}

\begin{document} 

\maketitle

\vspace{-11ex}

\begin{center}
\begin{tabular}{ccccc}%
    \large Juan Margalef-Bentabol${}^{1,3,4}$   &&  \large D. Leigh Herman${}^1$ &&   \large Ivan Booth${}^{1,2}$\\
    \normalsize 
    \href{mailto:juanmargalef@mun.ca}{juanmargalef@mun.ca}  &&\href{mailto:dlherman@mun.ca}{dlherman@mun.ca}  &&\href{mailto:ibooth@mun.ca}{ibooth@mun.ca}
  \end{tabular}%
  
  \small ${}^{1}$ Department of Mathematics and Statistics, Memorial University of Newfoundland, NL Canada\\
   ${}^{2}$ Department of Physics and Physical Oceanography, Memorial University of Newfoundland, NL Canada\\
  ${}^{3}$ Departamento de Matemáticas, Universidad Carlos III de Madrid, Madrid, Spain\\
  ${}^{4}$ Grupo de Teorías de Campos y Física Estadística. Unidad Asociada IGM-UC3M y IEM-CSIC
\end{center}

\mbox{}\vspace{-3ex}

\begin{abstract}

\mbox{}\vspace{-4.5ex}

\noindent Understanding degrees of freedom in classical mechanics is fundamental to characterizing physical systems. Counting them is usually easy, especially if we can assign them a clear meaning. However, the precise definition of a degree of freedom is not usually presented in first-year physics courses since it requires mathematical knowledge only learned in more advanced courses. In this paper, we use a pedagogical approach motivated by simple but non-trivial mechanical examples to define degrees of freedom and configuration spaces. We highlight the role that topology plays in understanding these ideas.
\end{abstract}

\section{Introduction}
When first-year physics students start learning classical mechanics, they are introduced to the concept of degrees of freedom and how to count them, but not so often the precise definition. Some inquisitive students will try to search for  this definition themselves, for instance, by looking at Wikipedia \cite{wikipedia}, where they would find the following: 
\begin{quote}
    In physics, the degrees of freedom (DOF) of a mechanical system is the number of independent parameters that define its configuration or state.
\end{quote}
with no further references to pursue. This might leave the students a bit puzzled, as this is the definition of the ``number of DOFs''. An astute student might explore mathematically advanced websites like Wolfram MathWorld \cite{mathworld} or nLab \cite{nlab} for clarification. The former offers a nearly identical definition while the latter provides nothing more than a link back to Wikipedia. Thus, so far, we have no definition of what is an \emph{individual} DOF.

Of course, a physics student cannot be discouraged by a few unsuccessful attempts and would continue their quest by delving into renowned texts on classical mechanics. In Feymann's lectures \cite{Feynman}, DOFs are only mentioned in the context of statistical physics. In the introduction of the fantastic book \emph{Mechanics} \cite{sommerfeld2016mechanics}, the author A.~Sommerfeld wrote back in the 1940s that ``the concept of degree of freedom is not too well known''. He then devoted Section 7 to DOFs and virtual displacement with plenty of examples. However, no definition of DOF is provided. At this juncture, the student may turn to the influential Landau and Lifshitz series, specifically to volume 1 titled \emph{Mechanics} \cite{landau1976mechanics}, where on the very first page, the student discovers the following definition:
\begin{quote}
    The number of independent quantities which must be specified in order to define uniquely the position of any system is called the number of degrees of freedom [..] Any $s$ quantities $q_1,q_2,\ldots,q_s$ which completely define the position of a system with $s$ degrees of freedom are called generalised co-ordinates of the system.
\end{quote}
As the previous definitions, this one still defines the number of DOFs rather than the concept itself. However, it introduces a new ingredient that might spark the student's curiosity, the potential relationship between DOFs and coordinates (while also pondering, coordinates of what?). Seeking further clarity, the student turns to another widely acclaimed book, \emph{Classical Mechanics} \cite{goldstein2002classical}, by H.~Goldstein, to find the following: ``A system of $N$ particles, free from constraints, has $3N$ independent coordinates or degrees of freedom''. Based on this definition, the hard-working student tentatively infers that coordinates and DOFs are interchangeable. However, a lingering question persists: coordinates of what specific space? It is worth noting, for the sake of completeness, that if our dedicated student delved into the seminal book \emph{Mathematical Methods of Classical Mechanics} \cite{Arnold} by V.~Arnold, the first mention to DOFs they would encounter is a seemingly different definition that relies on differential equations:
\begin{quote}
A system with one DOF is a system described by one differential equation $\ddot{x} = f(x),\ x\in \R$. 
\end{quote}
As this interesting approach is likely unfamiliar to first-year students, we will refrain from delving into it in this paper. Instead, we will focus on the static perspective, specifically studying simple mechanical systems to examine the relationship between coordinates and DOFs that our diligent students have speculated about. We will provide a formal definition of DOFs and introduce the concept of configuration space. We will see how the topology of this space plays a vital role, and we will explore more complex examples to highlight its significance.

\section{Degrees of freedom}
\subsection{Definitions and simple examples}

Let us begin with a system formed by a train (taken to be a point particle) on an infinite straight railroad. Clearly, the train's position can be fully described by one number. The obvious choice is the distance $d$ to a fixed origin, but it is not the only choice. For example, we could use $x=e^d$ or more complicated expressions as long as a unique number (\emph{coordinate}) is associated with each location. 
Most often, the choice of coordinate would be made to render the dynamics as simple as possible, but from a static perspective, all choices are equally good. 
Since only one number is required, this system is said to have a single degree of freedom (DOF). 

The position of the end-point of a planar pendulum can also be described with just one number, for example, the angle $\theta$ between the pendulum and the vertical axis (although, again, less obvious but equivalent choices are possible). Much like the train example, the planar pendulum has one DOF, but we can see that these DOFs are not completely equivalent. To return to the starting point after advancing in any direction, the train has to retrace its path over the track it previously traversed. However, the planar pendulum can also return to its starting point by continuing in the same direction to complete a full rotation. Indeed, $\theta$ and $\theta+2\pi$ represent the same configuration, corresponding to gluing the endpoints of the interval $[0,2\pi]$ to form a circle. Both examples have one DOF, which means that at each point, the object can move only in one direction. However, as we have mentioned before, they are globally very different. To capture the difference, we need to study all possible positions of the system at once, which leads to the following definition.

\begin{definition}
The configuration space (C-space for short) is the set formed by all possible positions of the system (also known as system configurations).\label{cspace}
\end{definition}



The C-spaces of our previous examples are the real line, $\mathbb{R}$, for the railroad and a circle, $\S^1$, for the pendulum. Mathematically, these are both examples of \emph{differentiable manifolds}. Roughly speaking, a manifold generalizes the idea of surfaces to any dimension. More formally: 

\begin{definition}\label{def:manifold}
An $n$-dimensional differentiable manifold is a space for which there is a neighbourhood around each point that looks like an open region of $\mathbb{R}^n$ (this is the ``manifold'' part). In particular, if one sets a coordinate system around any point, one can do calculus in the standard way using those coordinates (this is the ``differentiable'' part).
\end{definition}

A coordinate system can be thought of as a machine (a function) that picks a point of the manifold and assigns to it a unique label (an array of $n$ numbers).  The machine, together with the region it can label, is known as \emph{coordinate patch}. However, for an arbitrary manifold, a single machine is generally insufficient to cover all points. 

\begin{figure}\centering
\includegraphics[width=\textwidth]{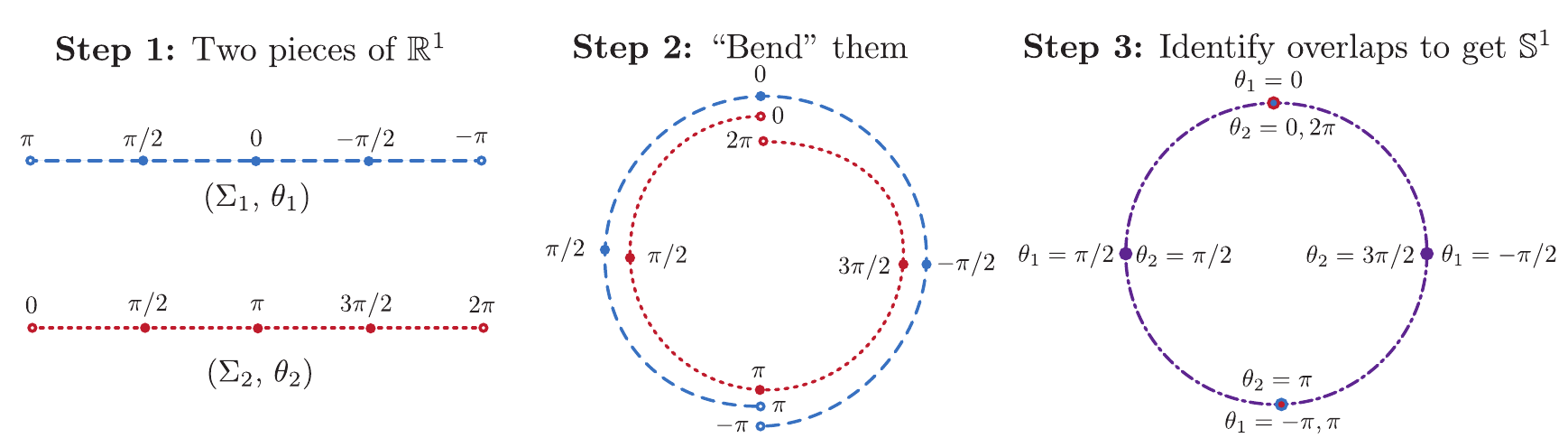}
\caption{Construction of a circle from two open pieces of $\mathbb{R}^1$ parameterized by coordinates $\theta_1$ and $\theta_2$ respectively. Open circles indicate that a point bounds but is not in the interval. In Step 3, the circle is colored purple to indicate the overlap of red and blue. With the identification, all points on the circle are in both coordinate systems except the ``north pole'', which is only in $\Sigma_1$, and the ``south pole'', which is only in $\Sigma_2$. }
\label{circle}
\end{figure}

The circle is the easiest example for which we can see that a single coordinate patch is not enough. 
We can understand the problem by considering a simple attempt to construct a coordinate
system: labelling every point on the circle as an angle. But, as is well-known, angle labels ``wrap-around'' the circle and so are only defined up to the addition of integer multiples of $2\pi$. Then, this is not a good coordinate: the labels returned by our machines should be unique. To get a unique label while covering as much of the circle as possible, we could restrict the available angles to the interval $(-\pi,\pi)$ (recall that we need open intervals) at the expense of missing one point (the ``south pole''). To cover that, a second coordinate system (another machine) is needed. For example, we could take the interval $(0,2\pi)$ to cover the missing point (see Fig.~\ref{circle}). Note that the patches overlap: all points except the north and south poles can be labelled in either coordinate system. 


From our previous discussion, a manifold is simply a set of patches of $\mathbb{R}^n$ that are nicely glued together. Thus, it looks locally like $\mathbb{R}^n$, but globally it might not. This difference is captured by the \emph{topology}, often described as ``rubber sheet geometry''. Topological properties are those that are unchanged if a manifold is stretched, squashed or otherwise deformed (ripping and gluing are not allowed). Returning to our examples of C-spaces, we see that $\mathbb{R}^1$ does not have the same topology as $\mathbb{S}^1$. As already noted, if you travel in one direction on $\mathbb{S}^1$, you will eventually return to where you started, but this is not true for $\mathbb{R}^1$. Neither of these conclusions depends on deformations of the manifold, and so whether or not a manifold loops around connecting to itself is a topological property. In our circle example, this looping was also the essential obstacle to covering $\mathbb{S}^1$ with a single coordinate patch.

\begin{figure}[ht!]
    \centering
    \includegraphics[clip,trim=15ex 2ex 11ex 2ex,height=.145\linewidth]{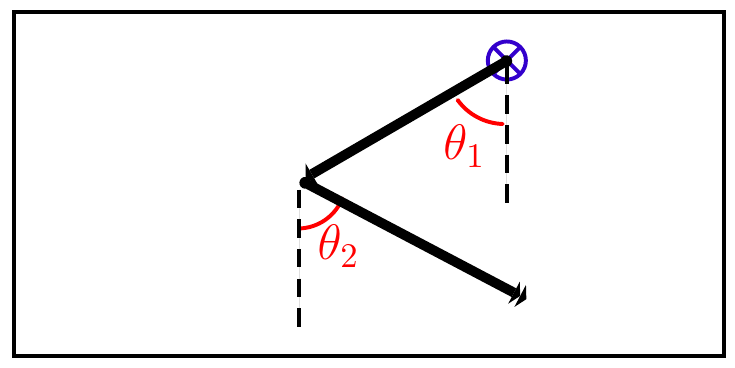}\hfill%
\includegraphics[width=.23\linewidth]{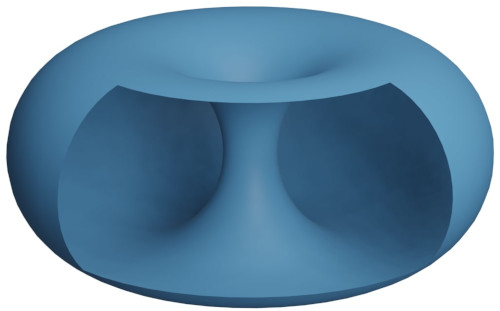}\hfill%
\includegraphics[clip,trim=2ex 2ex 3ex 2ex,height=.145\linewidth]{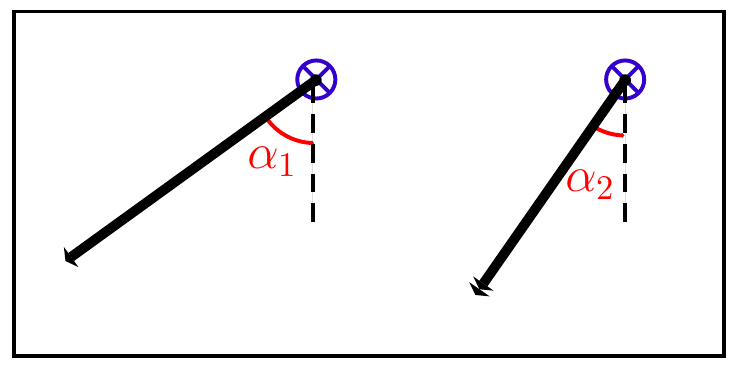}%
    \caption{The configuration spaces of the double planar pendulum (left) and two independent pendulums are both torii (center).}
    \label{double-double}
\end{figure}

With these ideas in mind, we next consider a double planar pendulum (left of Fig.~\ref{double-double}). The position of each pendulum is independent, so we need two angles: $\theta_1\in\S^1$ between the first pendulum and the vertical axis, and $\theta_2\in\S^1$ between the second pendulum and the vertical axis. A given configuration of the system is fully described by the numbers $(\theta_1,\theta_2)$, so the system has two DOFs. Each of these corresponds to a circle and so the C-space is $\S^1\times \S^1$, i.e. a torus. 

\begin{figure}[!ht]\centering
	\includegraphics[width=.88\textwidth]{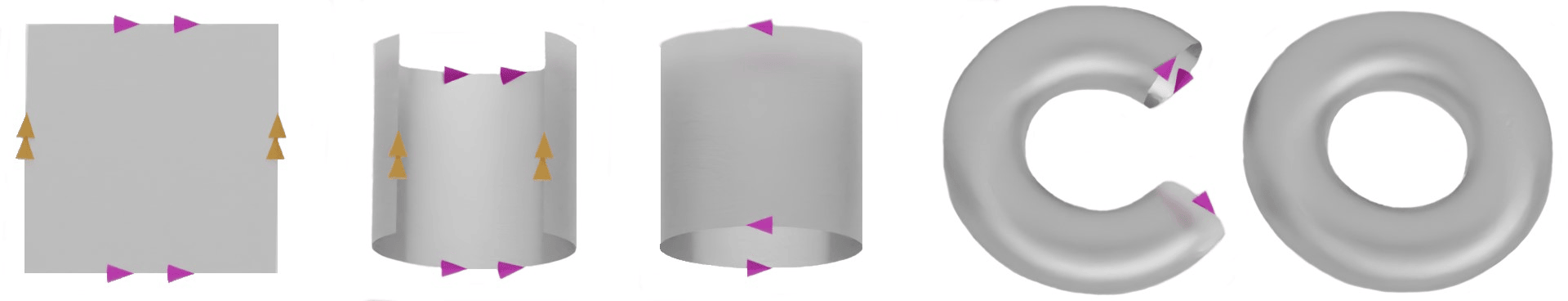}
	\caption{Gluing process to obtain a torus.}
	\label{fig:GlueTorus}
\end{figure}

We saw that the C-space of the pendulum is the circle, which can be obtained by gluing the endpoints of the interval $[0,2\pi]$. Analogously, the C-space of the double pendulum can be obtained by gluing the sides of the square $[0,2\pi]\times[0,2\pi]$,  which leads, as Fig.~\ref{fig:GlueTorus} shows, to a torus. Note that, as for the circle, a single coordinate patch is not enough to cover the torus.

Notice that if we consider the system formed by two independent pendulums (right of Fig.~\ref{double-double}), we again need two angles $(\alpha_1,\alpha_2)$ to fully describe the system, and we obtain that the C-space is also a torus (even though the dynamics of the
two systems are very different). %

From these and other standard mechanical examples, the reader may get the impression that the \mbox{C-space} is always a differentiable manifold. Surprisingly, as we will see in the next section, this is not always the case. However, in most of the common examples, the C-space is indeed a manifold; 
the C-space can be locally parameterized as a piece of $\mathbb{R}^n$. That is, we can choose a set of coordinates (remember that there are many equivalent choices) around a point by assigning labels (numbers) to its nearby points. By local, we mean that such labelling might not cover the whole manifold. With this in mind, we are ready to finally define what a degree of freedom is: 

\begin{definition}\label{Def:DOF}
A degree of freedom is a local coordinate on the C-space of the mechanical system.
\end{definition}

As for every manifold, DOFs (coordinates) enumerate the ways one can independently move through the C-space but are only valid locally. Moreover, the number of DOFs is simply the dimension of the C-space (a definition our student would have discovered in \cite{Arnold} had they been more patient to get to chapter 4!). In particular, for finite-dimensional C-spaces, $\#\mathrm{DOFs}=n-r$ where $n$ and $r$ are the numbers of variables and constraints \cite{goldstein2002classical,Arnold}. However, the fact that the DOFs are just coordinates (something known since Lagrange started to study C-spaces) means that they do not always have a direct physical or geometric meaning. Thus, asking ``what'' is a given DOF or ``where'' it is located might be trickier than expected. This will often be the case if one is studying a problem with a more complicated geometry
and using generalized coordinates \cite{goldstein2002classical}.

For example, let us consider the movement of a particle moving on an elliptical track
\begin{equation}
    \frac{x^2}{a^2}+ \frac{x^2}{b^2} = 1 \; . 
\end{equation}
We have two coordinates, $x$ and $y$, and one constraint (the above equation), which leads to one DOF. The most convenient parameter to describe this system is $\lambda$, for which
\begin{eqnarray}
    x = a \cos \lambda \; \; \mbox{ and } \; \; y = b \sin \lambda \; . 
\end{eqnarray}
 However, it does not have an immediate physical or geometric meaning. It is neither the angle between $(x,y)$ and one of the coordinate axes nor a measure of distance along the curve. These are:
 \begin{equation}
\theta = \arctan \left(\frac{b}{a} \tan \lambda \right)     \qquad \mbox{and} \qquad
s = a \int_{t=0}^\lambda \sqrt{1-\frac{b^2}{a^2}\sin^2 \! t}  \, \mathrm{d} t
 \end{equation}
where the integral does not have a closed-form solution but is well-studied as an elliptic integral of the second kind. In this case, either $\lambda$, $\theta$ or $s$ could be used as the DOF since they are equally valid (and equivalent), but $\lambda$, the least physical, is the most convenient.

\subsection{Tricky example}

\begin{figure}[!ht]
     \centering  
     \includegraphics[clip,trim={.25cm .3cm 1.5cm .6cm},width=.46\textwidth]{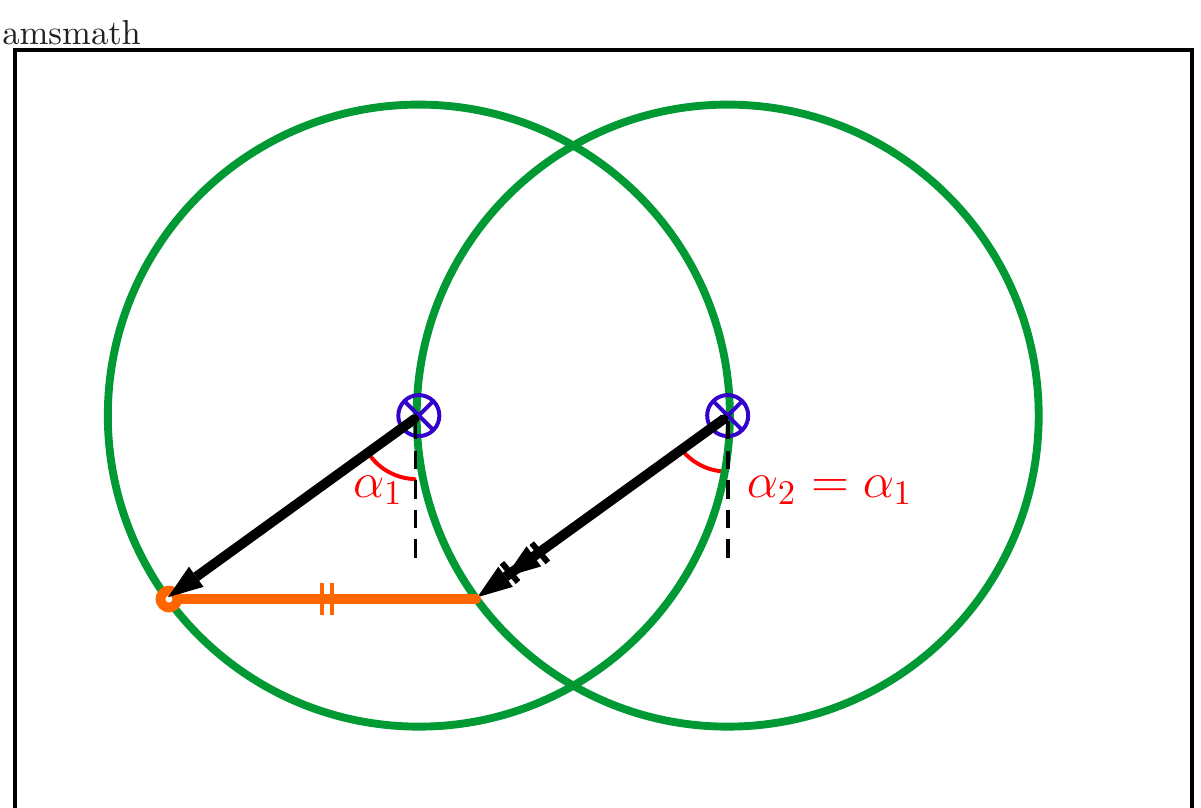}
     \caption{System formed by two pendulums joined by the endpoints with a rigid rod. The circles represent the track that the endpoints follow.}\label{Fig. double pendulum joined circles}
\end{figure}

This next example came up years ago in conversations about DOFs between one of the authors (J.M-B) and Fernando Barbero. Consider two pendulums of length $\ell$ with pivots at $q_1=(0,0)$ and $q_2=(\ell,0)$, and with their endpoints $p_1$ and $p_2$ joined by a rigid rod of length $\ell$ (see Fig. \ref{Fig. double pendulum joined circles}).

Counting DOFs is easy: four variables (two coordinates for each $p_i$) and three constraints (one for each rod) lead to one DOF. But, what is the C-space of this system? The first guess is that this system is equivalent to a single pendulum since both pendulums seem locked (the angles $\alpha_i$ of the $i$-th pendulum with the vertical axis are equal). Hence, the naive guess is that the C-space is $\S^1$. 

\begin{figure}[!ht]
     \centering  
     \includegraphics[clip,trim={.4cm .3cm 3cm .6cm},width=.5\textwidth]{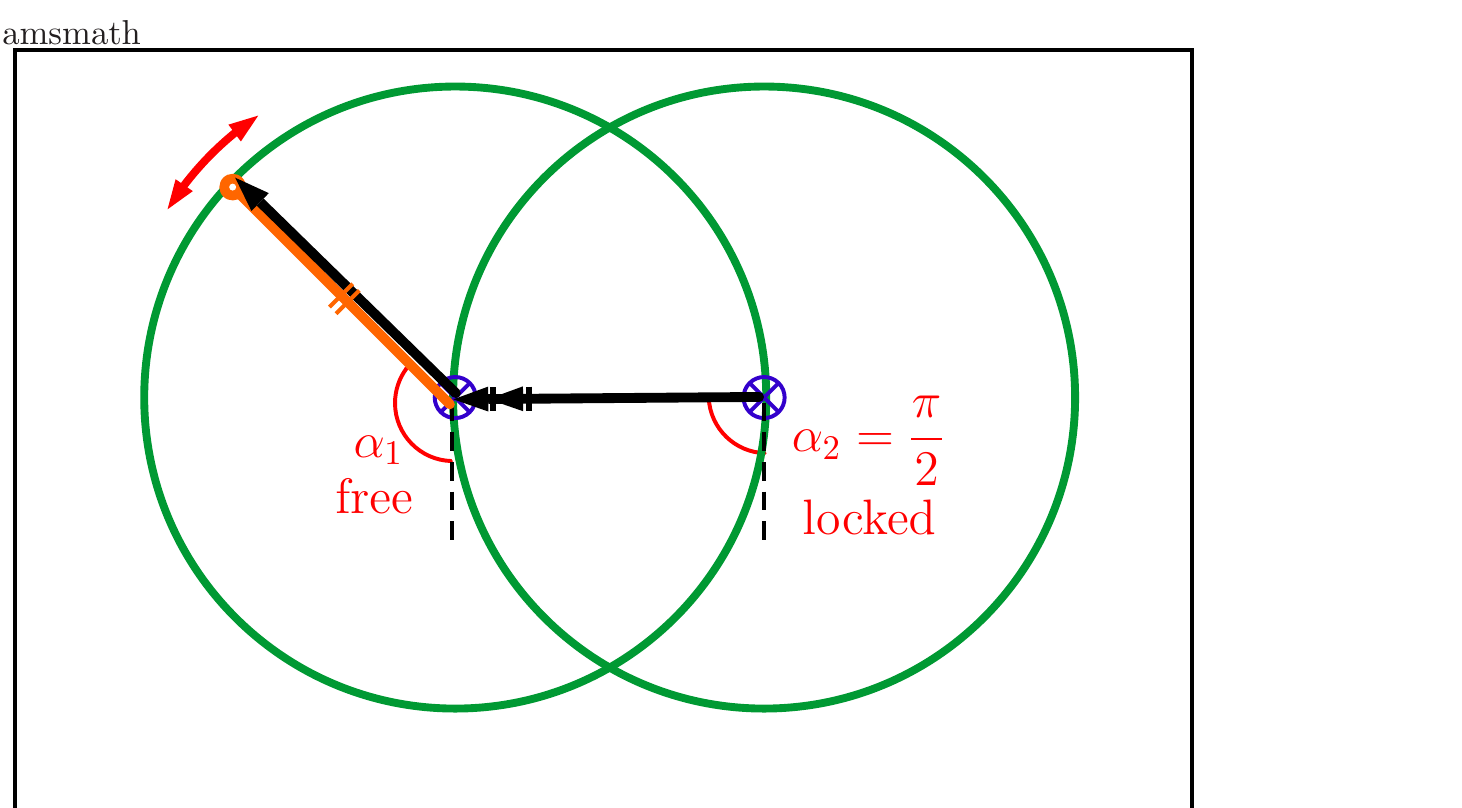}\includegraphics[clip,trim={.4cm .3cm .5cm .6cm},width=.5\textwidth]{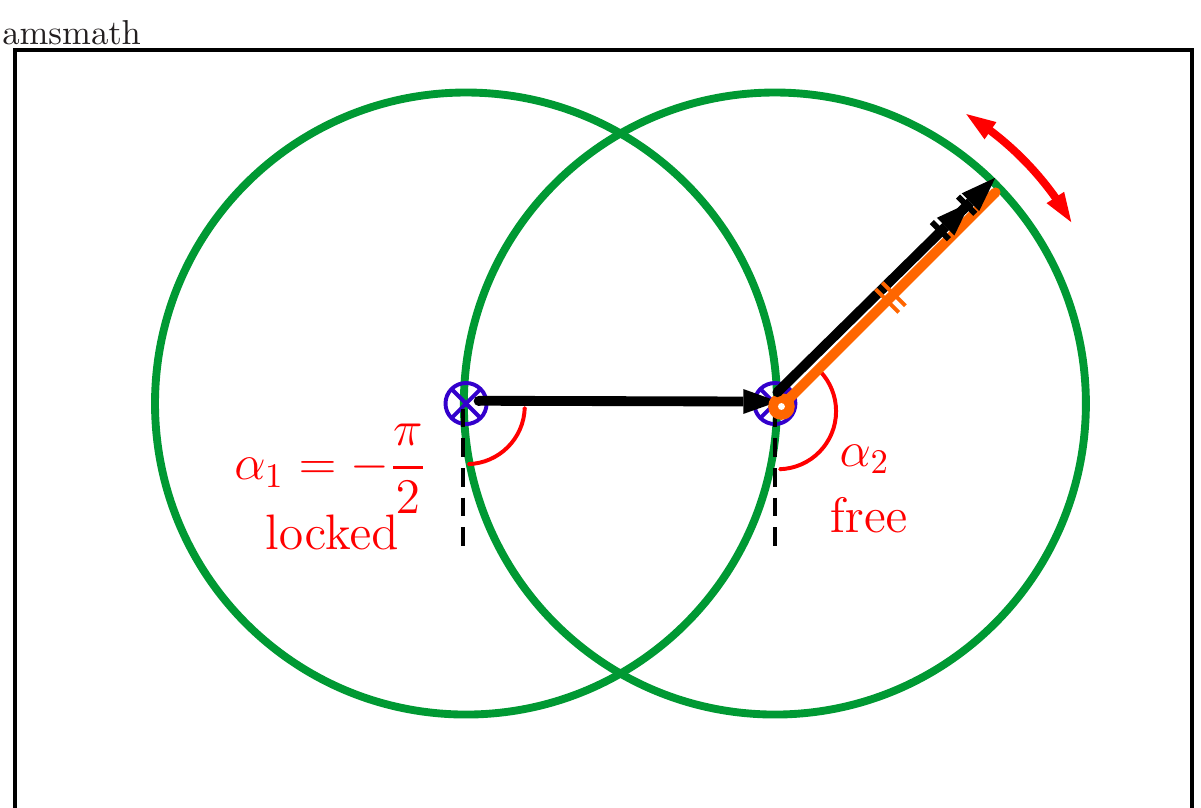}
     \caption{The system with one of the pendulums locked.}\label{Fig. locked systems}
\end{figure}

\begin{figure}[!ht]
     \centering  
     \includegraphics[clip,trim={.4cm .3cm 3cm .6cm},width=.44\textwidth]{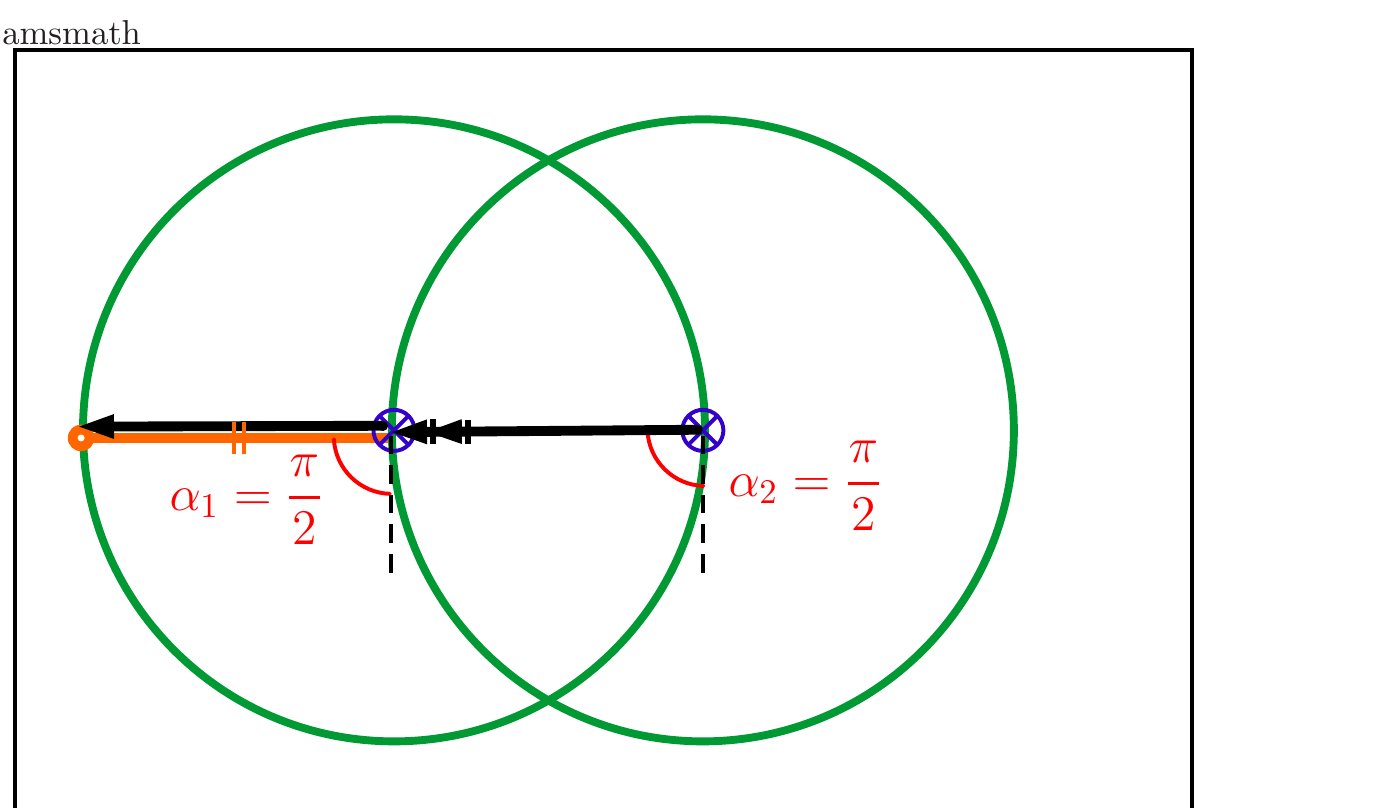}\includegraphics[clip,trim={.4cm .3cm 3cm .6cm},width=.44\textwidth]{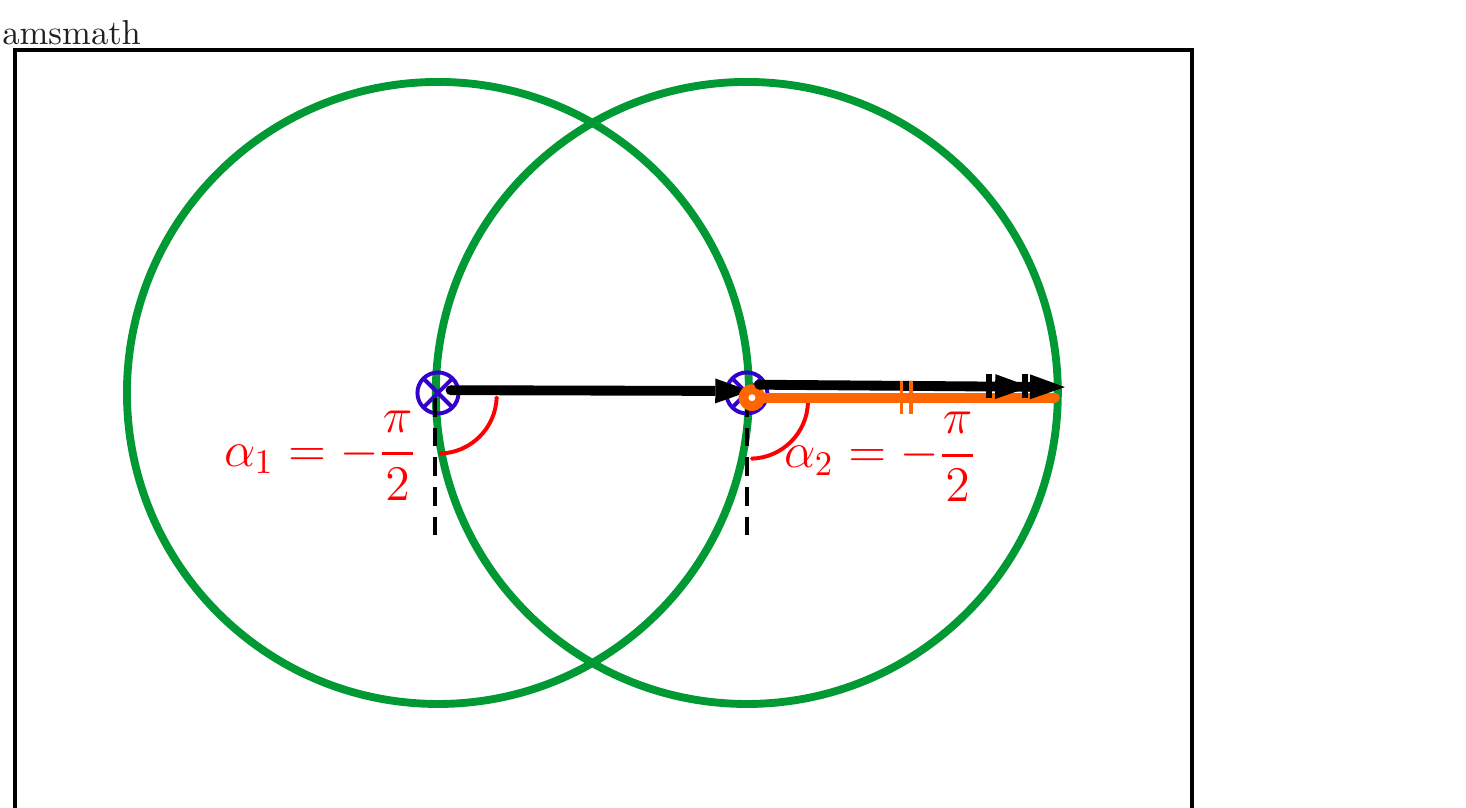} \\
     \includegraphics[clip,trim={.4cm .3cm 3cm .6cm},width=.44\textwidth]{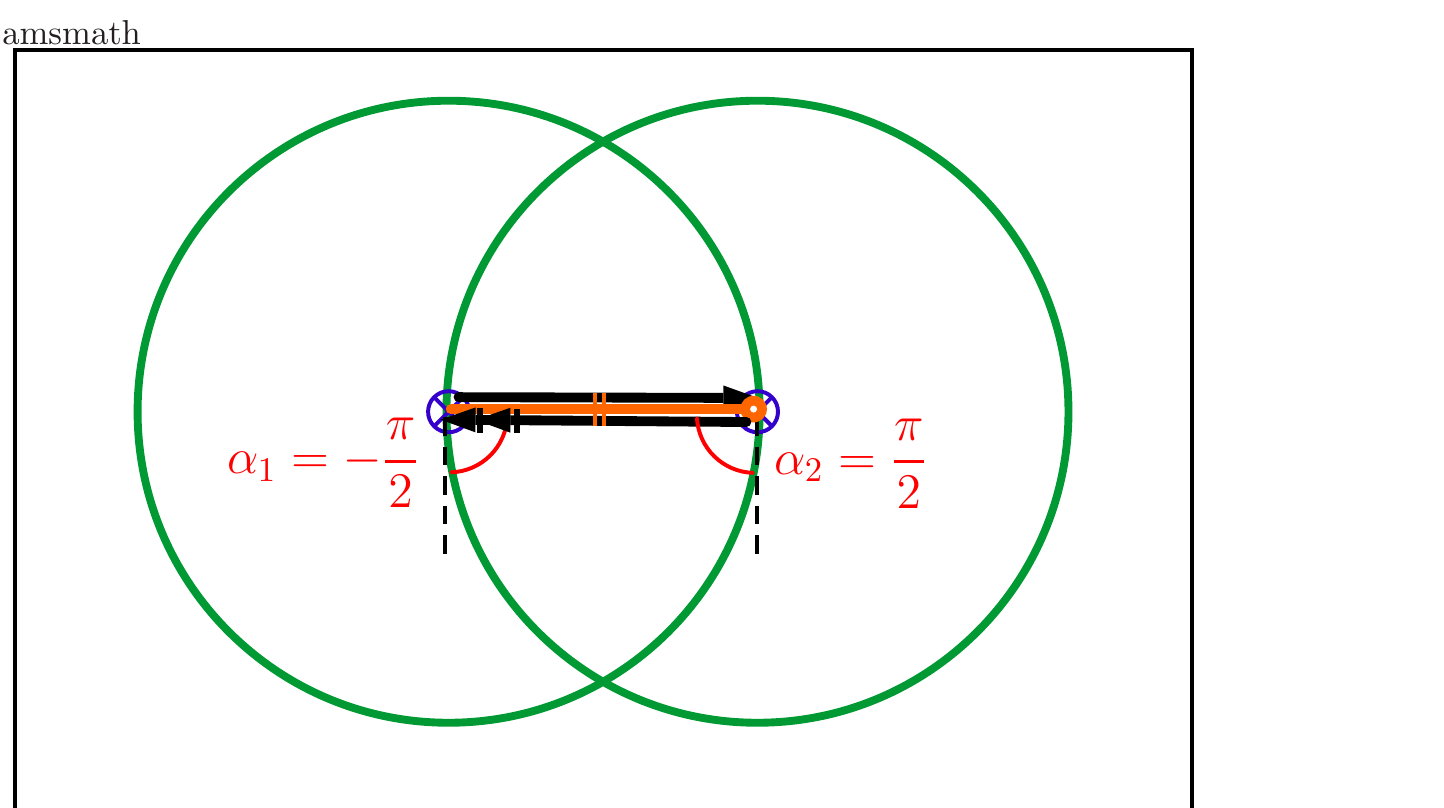}
     \caption{The configurations that lie in the intersection of two $\mathcal{C}_i$.}\label{Fig. configuration intersection}
\end{figure}

However, there is a little caveat: the pendulums can actually be moved separately in some cases. Indeed, if we place the system on the horizontal axis (i.e., $\alpha_1=\alpha_2=\pm\pi/2$, see Fig.~\ref{Fig. configuration intersection}), we can lock the pendulum joining the centers and move the other one freely! That behaviour is portrayed in Fig.~\ref{Fig. locked systems}.

We see that the C-space can be parametrized by
\begin{equation}
\mathcal{C}=\mathcal{C}_1\cup\mathcal{C}_2\cup\mathcal{C}_3\quad\text{where}\quad\begin{array}{l}
    \mathcal{C}_1=\left\{(\alpha_1,\alpha_2)\in\S^1\times\S^1\ \text{ such that } \alpha_1=\alpha_2\right\}\\[2ex]
    \mathcal{C}_2=\left\{(\alpha_1,\alpha_2)\in\S^1\times\S^1\ \text{ such that } \quad \begin{array}{|l}\!\alpha_1\text{ arbitrary}\\ \!\alpha_2=\frac{\pi}{2}\end{array}\right\}\\[3.5ex]
    \mathcal{C}_3=\left\{(\alpha_1,\alpha_2)\in\S^1\times\S^1\ \text{ such that }  \begin{array}{|l}\!\alpha_1=-\frac{\pi}{2}\\ \!\alpha_2\text{ arbitrary}\end{array}\right\}
\end{array}
\end{equation}
$\mathcal{C}_1$ parametrizes the configurations for which both pendulums are moved together (Fig.~\ref{Fig. double pendulum joined circles}); $\mathcal{C}_2$ the configurations for which the second pendulum is locked and the first is free (left image of Fig.~\ref{Fig. locked systems}); and $\mathcal{C}_3$ the configurations for which the first pendulum is locked and the second is free (right image of Fig.~\ref{Fig. locked systems}). Each $\mathcal{C}_i$ is equivalent to a circle, but they are clearly not disjoint. Indeed, $(\pi/2,\pi/2)\in\mathcal{C}_1\cap\mathcal{C}_2$, $(-\pi/2,-\pi/2)\in\mathcal{C}_1\cap\mathcal{C}_3$, and $(-\pi/2,\pi/2)\in\mathcal{C}_2\cap\mathcal{C}_3$ (see Fig. \ref{Fig. configuration intersection}).

\begin{figure}[!ht]
     \centering  \includegraphics[clip,trim={.1cm .1cm 3.7cm .1cm},page=2,width=.51\textwidth]{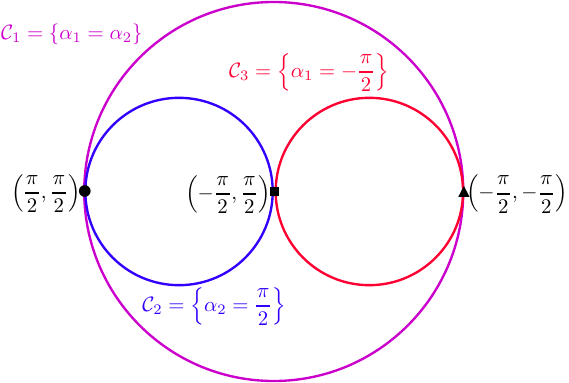}
     \caption{C-space of the tricky example with the three intersection points labelled with a circle, a square, and a triangle.}
         \label{fig:Cspacetrickyexample}
\end{figure}

 The C-space is formed by three circles with an intersection point between any pair of them (see Fig.~\ref{fig:Cspacetrickyexample}).

Away from the three intersection points, there is only one direction to evolve the system, so the C-space is clearly one-dimensional. However, at the intersection points, something more complicated happens: there is a choice from two paths along which the system can evolve \footnote{Though, unlike a system with two true DOFs, it cannot evolve in a combination of those two directions. It has to choose one or the other.}. 
In some sense, the C-space has an extra \emph{discrete} freedom at each of those three points in choosing which path to follow, but this is not a DOF. Indeed, the presence of these points means that the C-space is not a manifold. Mathematically, at each of these three points (and only there), no matter how far we zoom in, we always see two intersecting lines, not the single that it would be if it was a one-dimensional manifold. Because of these three points, the C-space does not satisfy Definition \ref{def:manifold}, and we cannot use Definition~\ref{Def:DOF}.



\begin{figure}[!ht]
     \centering  \includegraphics[clip,trim={0 0 29ex 0},page=2,width=.55\textwidth]{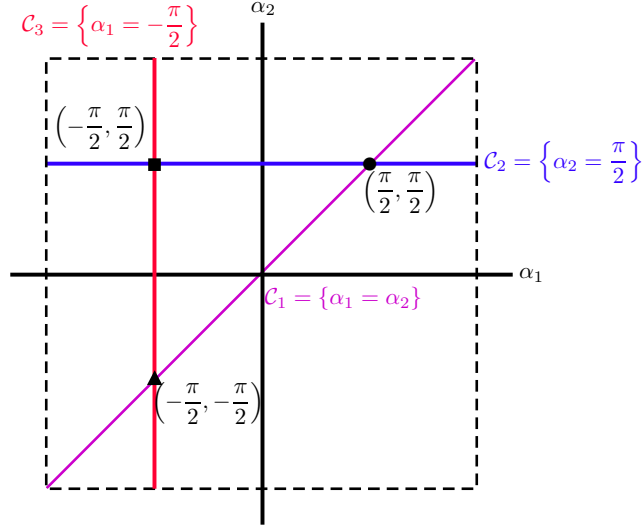} 
     \caption{Alternative visualization of the C-space of the tricky example. We see the same three circles as in Fig.~\ref{fig:Cspacetrickyexample} with the three intersection points labelled with a circle, a square, and a triangle.}
         \label{fig:restricteddoublependulum}
\end{figure}

Before ending this section, it is worth exploring another visual way to obtain the C-space: realizing that the system is a constrained double planar pendulum. Indeed, we have the double pendulum formed by $q_1$-$p_1$-$p_2$ with the additional constraint that the distance from $p_2$ to $q_2$ is $\ell$. This means that the C-space is a subset of the C-space of an unconstrained double pendulum, i.e.,  of a torus. Using the square representation of a torus (introduced in Fig.~\ref{fig:GlueTorus}), we see that $\mathcal{C}_1$ corresponds to the diagonal $\{\alpha_1=\alpha_2\}$, $\mathcal{C}_2$ the horizontal line  $\{\alpha_2=\pi/2\}$, and $\mathcal{C}_3$ the vertical line  $\{\alpha_1=-\pi/2\}$ as Fig.~\ref{fig:restricteddoublependulum} shows.

Now, if we perform the gluing process described in Fig.~\ref{fig:GlueTorus}, we obtain a torus with the embedded C-space formed by three circles. Readers who are familiar with knot theory might recognize that they are torus knots of degree $(1,1)$ for $\mathcal{C}_1$, $(1,0)$ for $\mathcal{C}_2$, and $(0,1)$ for $\mathcal{C}_3$ (a  knot over a torus has degree $(p,q)$ if it winds $q$ times around the ``vertical'' circle and $p$ times around its axis of rotational symmetry).

\begin{figure}[!ht]
     \centering
     \includegraphics[width=.55\textwidth]{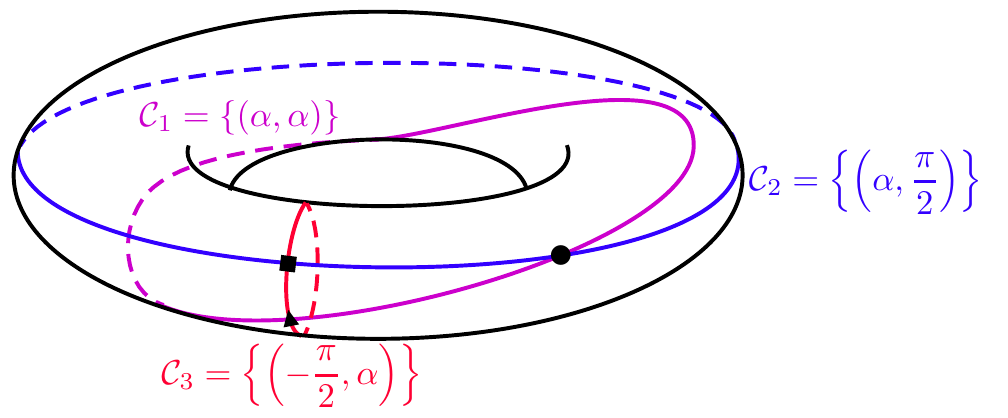}
     \caption{Alternative visualization of the C-space of the tricky example obtained after gluing the sides of Fig.~\ref{fig:restricteddoublependulum} following the process depicted in Fig.~\ref{fig:GlueTorus}.}
     \label{fig:restricteddoublependulum2} 
\end{figure}

This example shows the difficulty of placing, counting, and computing DOFs in some non-trivial examples. We see that $\alpha_1$ and $\alpha_2$ are both required to describe the system, even though the system has only one DOF. As we have seen, we have three cases: \begin{enumerate}
    \item If $\alpha_1$ is fixed at $\pi/2$, the freedom lies on $\alpha_2$.
    \item If  $\alpha_2$ is fixed, the freedom lies on $\alpha_1$.
    \item If none of these conditions holds, then $\alpha_1=\alpha_2$ and the freedom is ``shared'' between both angles.
\end{enumerate} 
The problem here stems, as we mentioned before, from the fact that the C-space of this example is not a manifold (see definition \ref{def:manifold}). 
The more advanced readers might find it interesting that the same difficulty in placing DOFs appears for gauge freedom, even for mechanical systems  \cite{prieto2015gauge}, and for boundary DOFs in field theories \cite{juarez2015quantization}.

\section{The importance of topology in C-spaces}\label{manifolds as config}

In this section, we focus on the second part of the title of this paper. Namely, we explore the importance of topology to properly understand the C\=/space of mechanical systems. Moreover, this section provides a nice motivation for introducing topology to physics students.

\subsection{Double planar pendulum}\label{subsection: double pendulum}
We found that the C-space of the double planar pendulum is a torus. In this section, we will rederive this fact using an approach that can be generalized to more complicated mechanical systems.  In what follows, we assume that the first pendulum has length $L$ and is attached to the origin, while the second has length $\ell<L$. We will focus on the endpoint $p$ of the second pendulum, where we place a pen to paint the largest possible circle $\S^1_{r^+}$ of radius $r^+:=L+\ell$ (max-circle). For that, we place both pendulums aligned and pointing in the same direction (straightened position, denoted by $\straight$, representing both pendulums pointing in the same direction). Analogously, the smallest circle around the origin that we can draw is $\S^1_{r^-}$ with $r^-:=L-\ell$ (min-circle), obtained when both pendulums are aligned and pointing in opposite directions (anti-straightened position, denoted by $\antistraight$, which represents the first pendulum pointing away from the center and the second one pointing back to the center).

\begin{figure}[h!]
\centering\includegraphics[width=.9\textwidth]{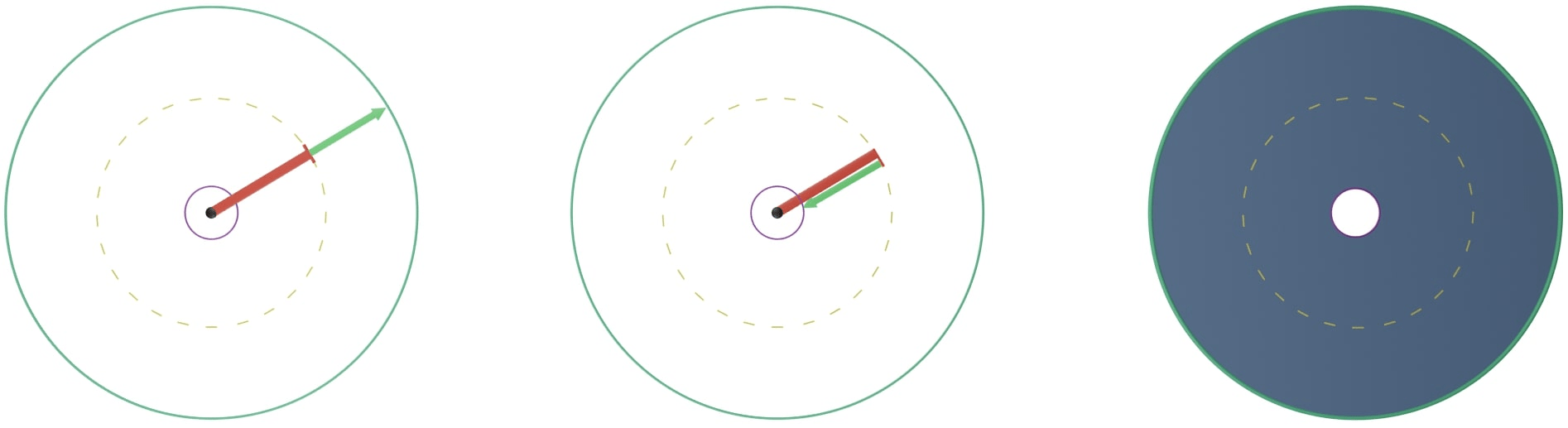}
    \caption{Straight configuration $\straight$ over the max-circle (left). Anti-straight configuration $\antistraight$ over the min circle (middle). On the right, we shade the region the pen can reach, i.e., between the max-circle and min-circle.}
    \label{fig:circles}
\end{figure}

\begin{figure}[!ht]
\centering\includegraphics[width=0.75\textwidth]{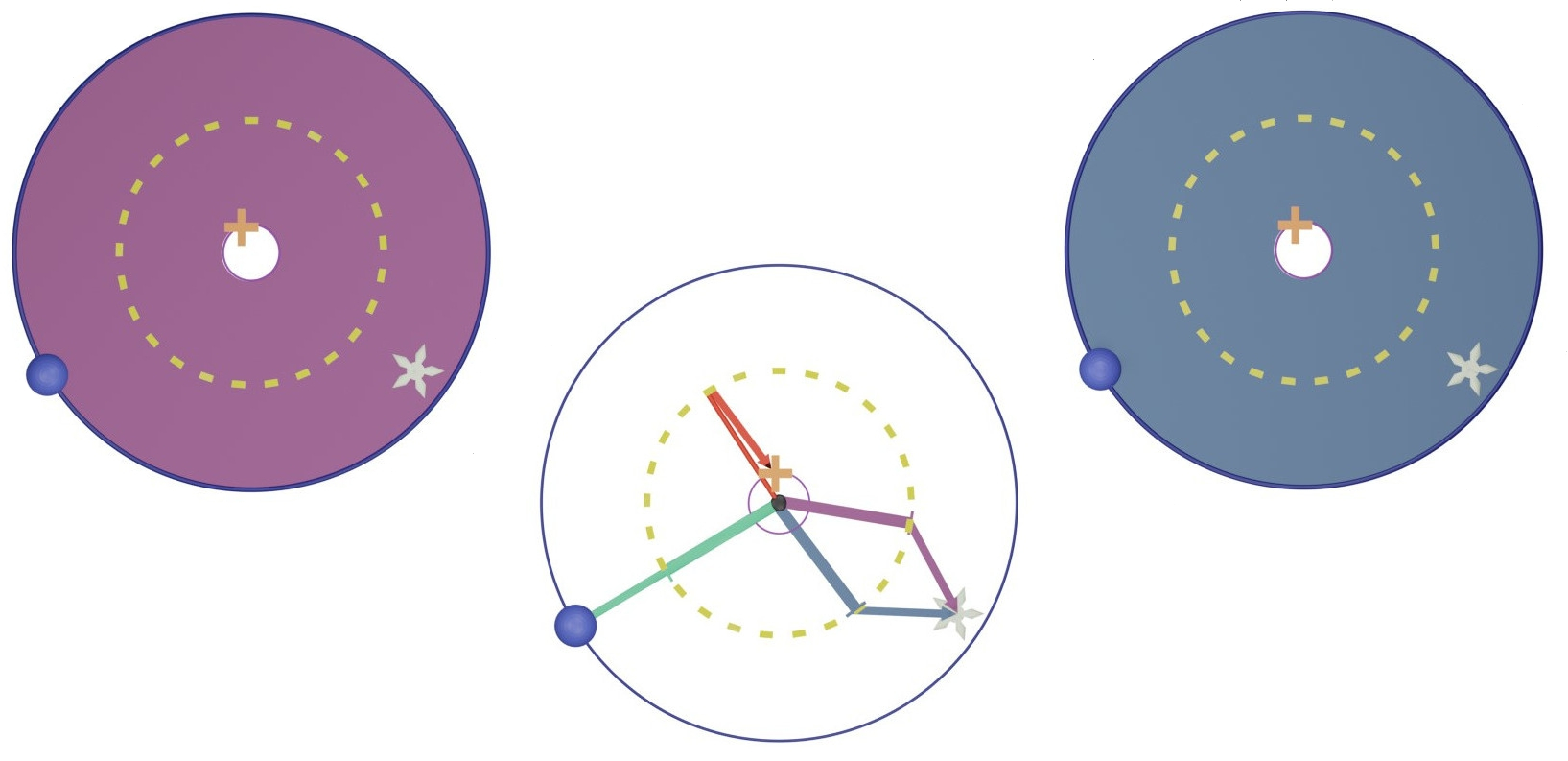}
    \caption{The C-space of the double pendulum on top: two copies of the annulus corresponding to the $\counterclock$ (left) and $\clock$ (right) configurations. Their boundaries are the limit case with the $\straight$ and $\antistraight$ configurations. On the bottom, we have the physical world where several configurations are drawn over it: one $\counterclock$ in purple, one $\clock$ in blue (both with equal endpoints), one $\straight$ in green, and one $\antistraight$ in red.}
    \label{fig:configurations}
\end{figure}

Since the pen at $p$ can also draw over the area in between the max-circle and min-circle, we can colour the interior of the annulus as well, which, in polar coordinates, is given by
\begin{equation}A(L,\ell):=\{(r,\theta)\in\R^+\times\S^1\ |\ r^-< r<r^+ \} \; .
\end{equation}
Then, $\mathscr{A}:=\S^1_{r^+}\cup A(L,\ell)\cup \S^1_{r^-}$ is the space of all points that $p$ reaches, so we will call it the \textit{reachable space} or \textit{R-space} (see Fig.~\ref{fig:circles}). Notice that this is not the full C-space since there are two configurations for every point of $A(L,\ell)$ 
as shown in Fig.~\ref{fig:configurations}. Namely, one when the pendulums are bent in the clockwise direction (that we denote as $\counterclock$) and the symmetric one with respect to the vector $\overline{Op}$ when they bend in the counterclockwise direction (that we denote as $\clock$).

\begin{figure}[!ht] 
    \hfill
    \begin{subfigure}[b]{0.2\textwidth}
        \includegraphics[width=.9\textwidth]{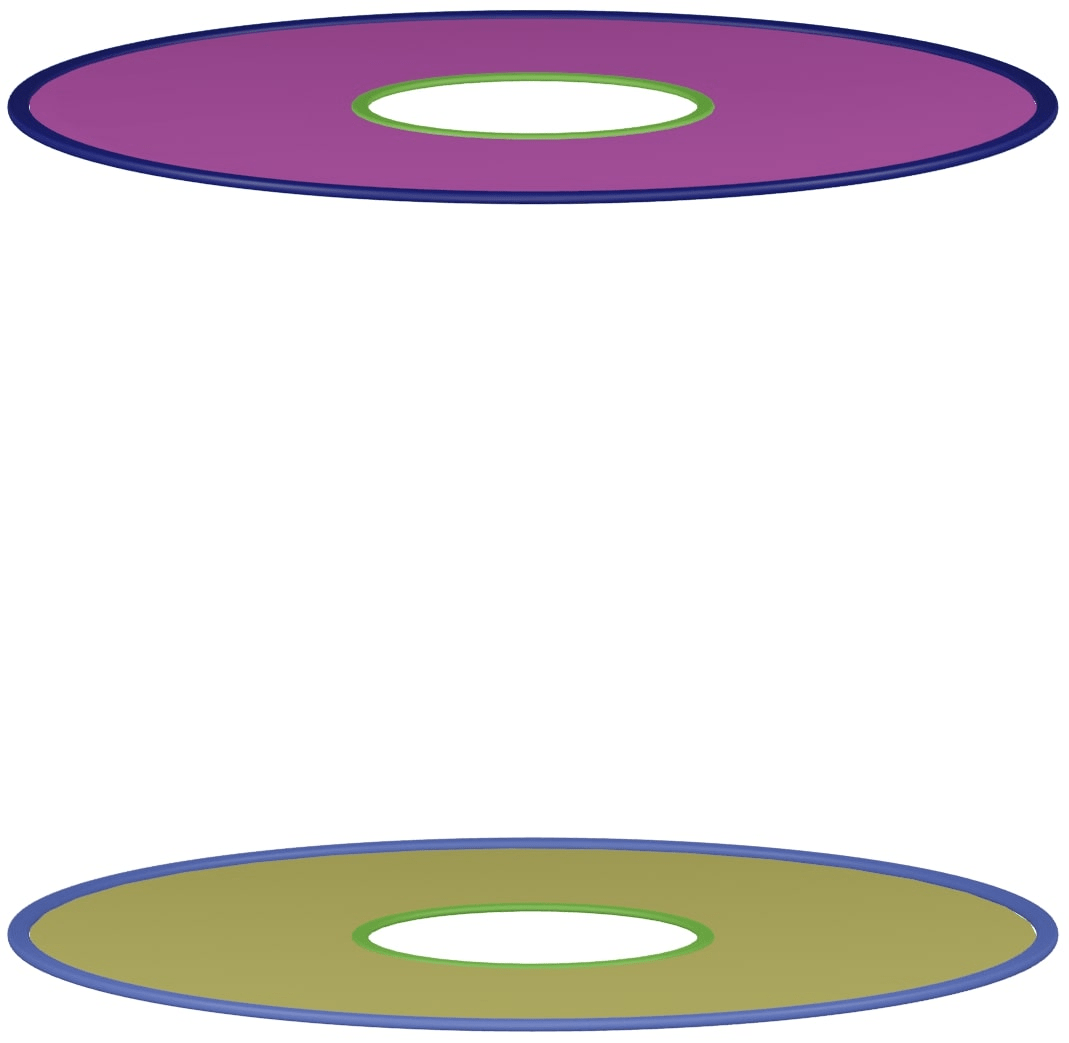}
    \end{subfigure}
    \hfill
    \begin{subfigure}[b]{0.25\textwidth}
        \includegraphics[width=.9\textwidth]{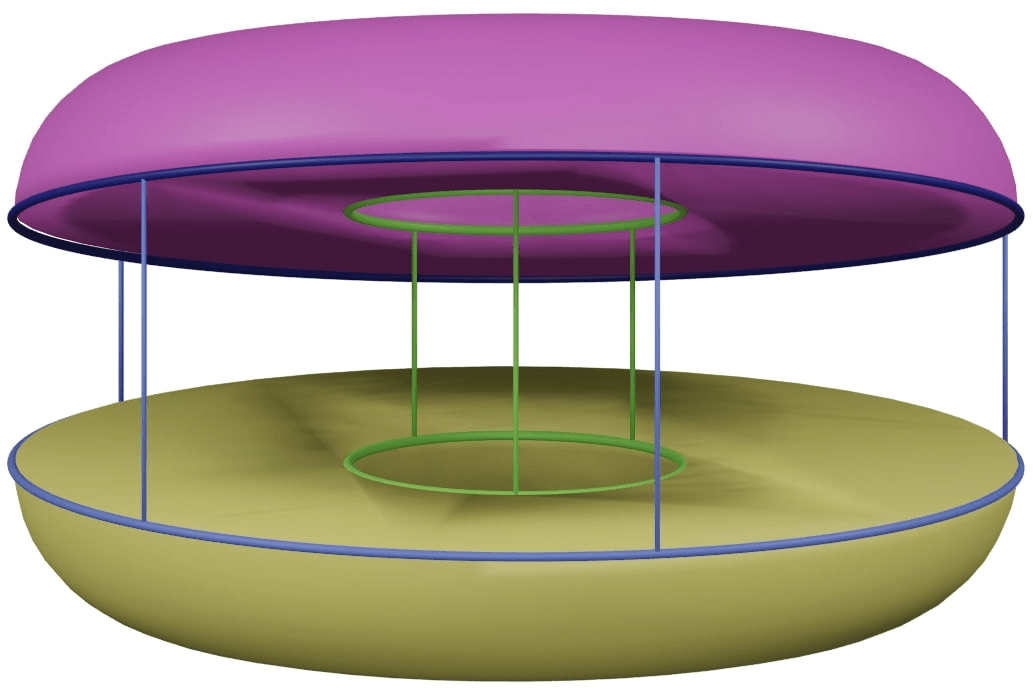}
    \end{subfigure}
    \hfill
    \begin{subfigure}[b]{0.25\textwidth}
        \includegraphics[width=.9\textwidth]{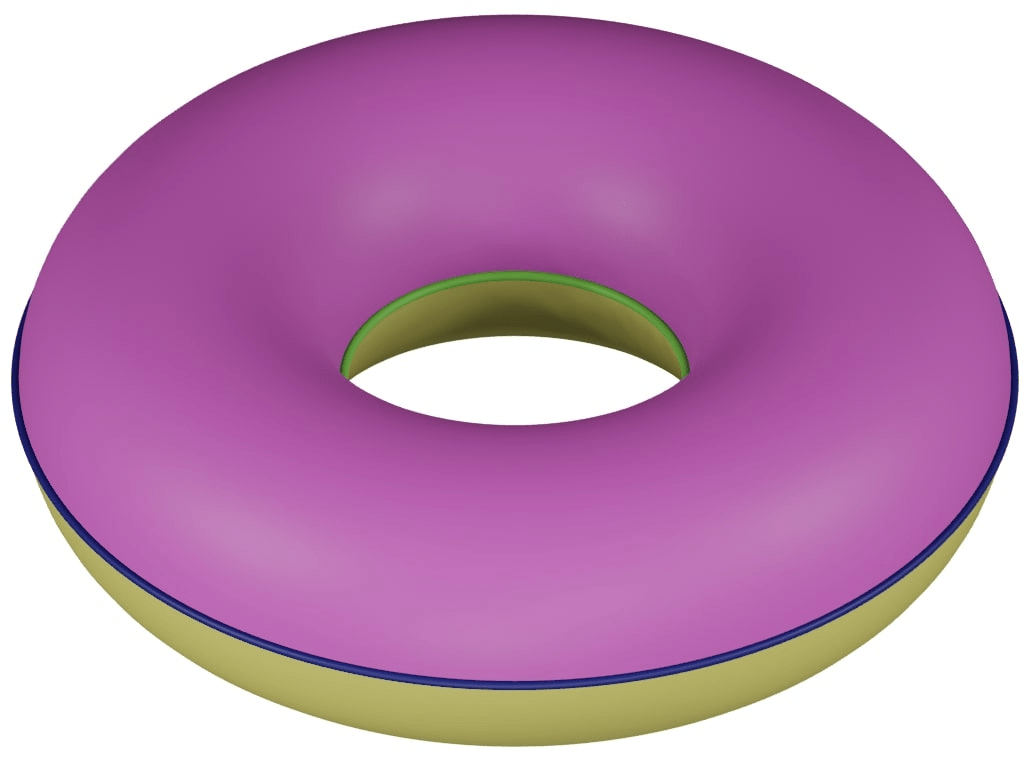}
    \end{subfigure}
    \hfill
    \caption{Two annuli with their boundaries identified give rise to a torus.}
    \label{fig:annulusToTorus}
\end{figure}

 This might lead to thinking that the C-space is two disjoint copies of $\mathscr{A}$, one corresponding to $\counterclock$ and the other to $\clock$. However, most readers probably have realized that the previous reasoning does not apply to the boundary circles. Indeed, the points of the max-circle $\S^1_{r^+}$ only have the $\straight$ configuration while the points of the min-circle $\S^1_{r^-}$ only have the $\antistraight$ configuration. In fact, these last two configurations are the limit of both $\clock$ and $\counterclock$ when we approach the corresponding boundary. The C-space is then obtained, as shown in Fig.~\ref{fig:annulusToTorus}, by taking two copies of $\mathscr{A}$ and identifying (gluing) the interior boundaries together with the same orientation and analogously for the exterior boundaries. In doing so, we obtain the expected torus.
  
\subsection{Linkages}\label{linkages}

\begin{figure}[!ht]
    \hspace{2em}
    \begin{subfigure}[b]{0.3\textwidth}
        \includegraphics[scale=.07]{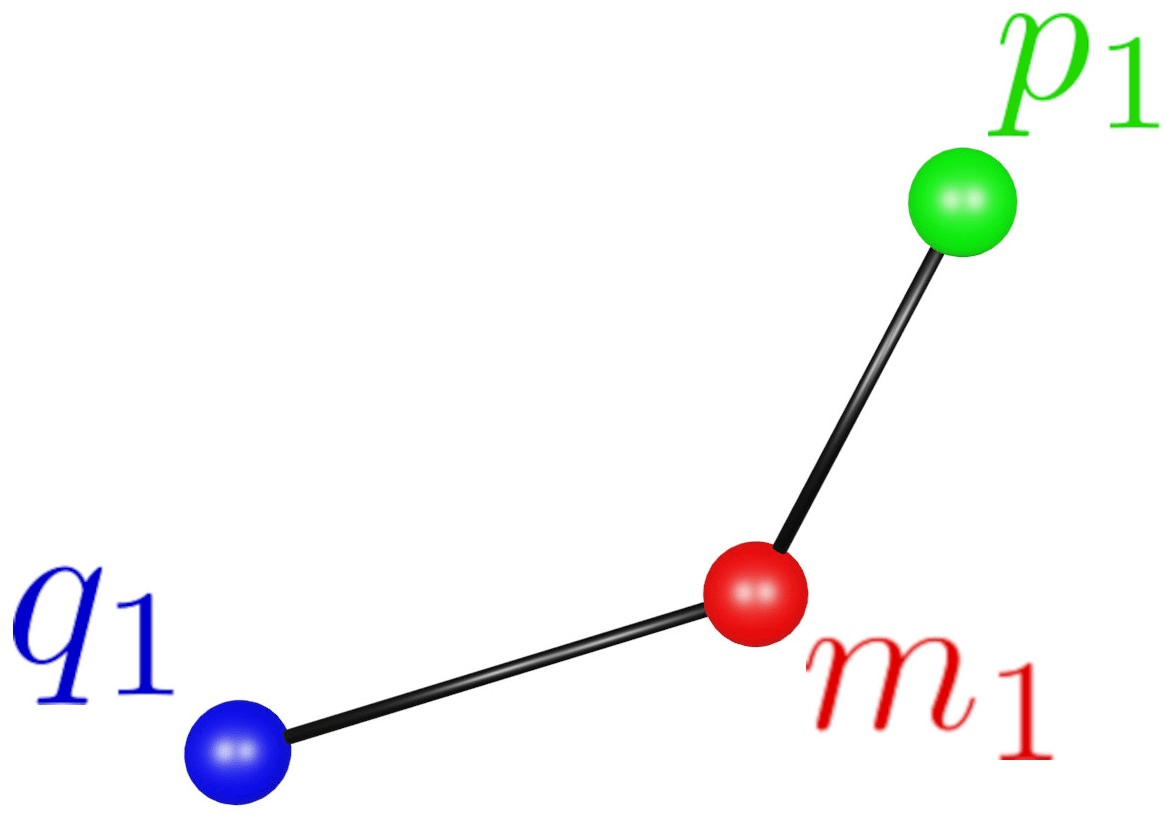}
    \end{subfigure}
    \hspace{-3em}
    \begin{subfigure}[b]{0.3\textwidth}
        \includegraphics[scale=.06]{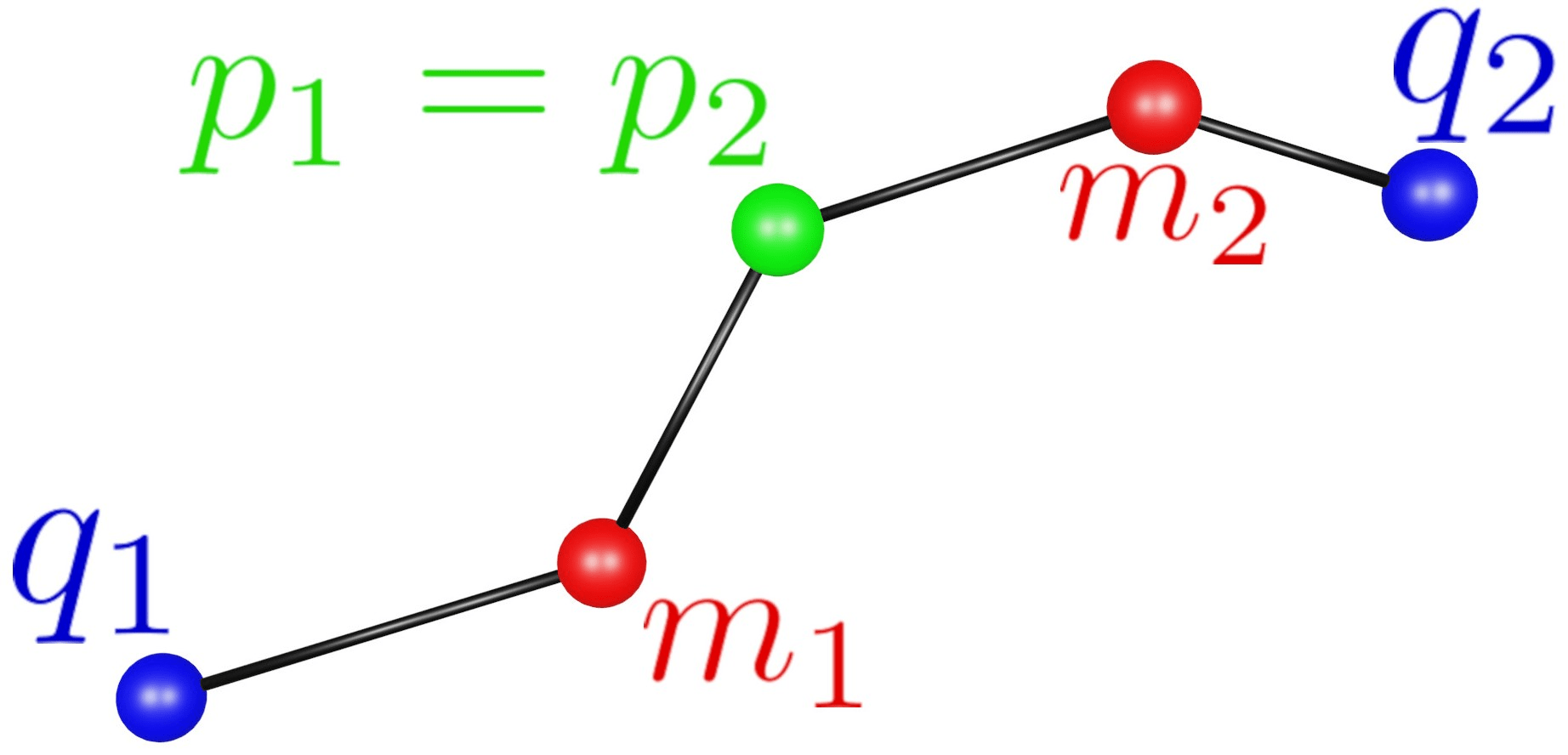}
    \end{subfigure}
    \hspace{1.5em}
    \begin{subfigure}[b]{0.3\textwidth}
        \includegraphics[scale=.068]{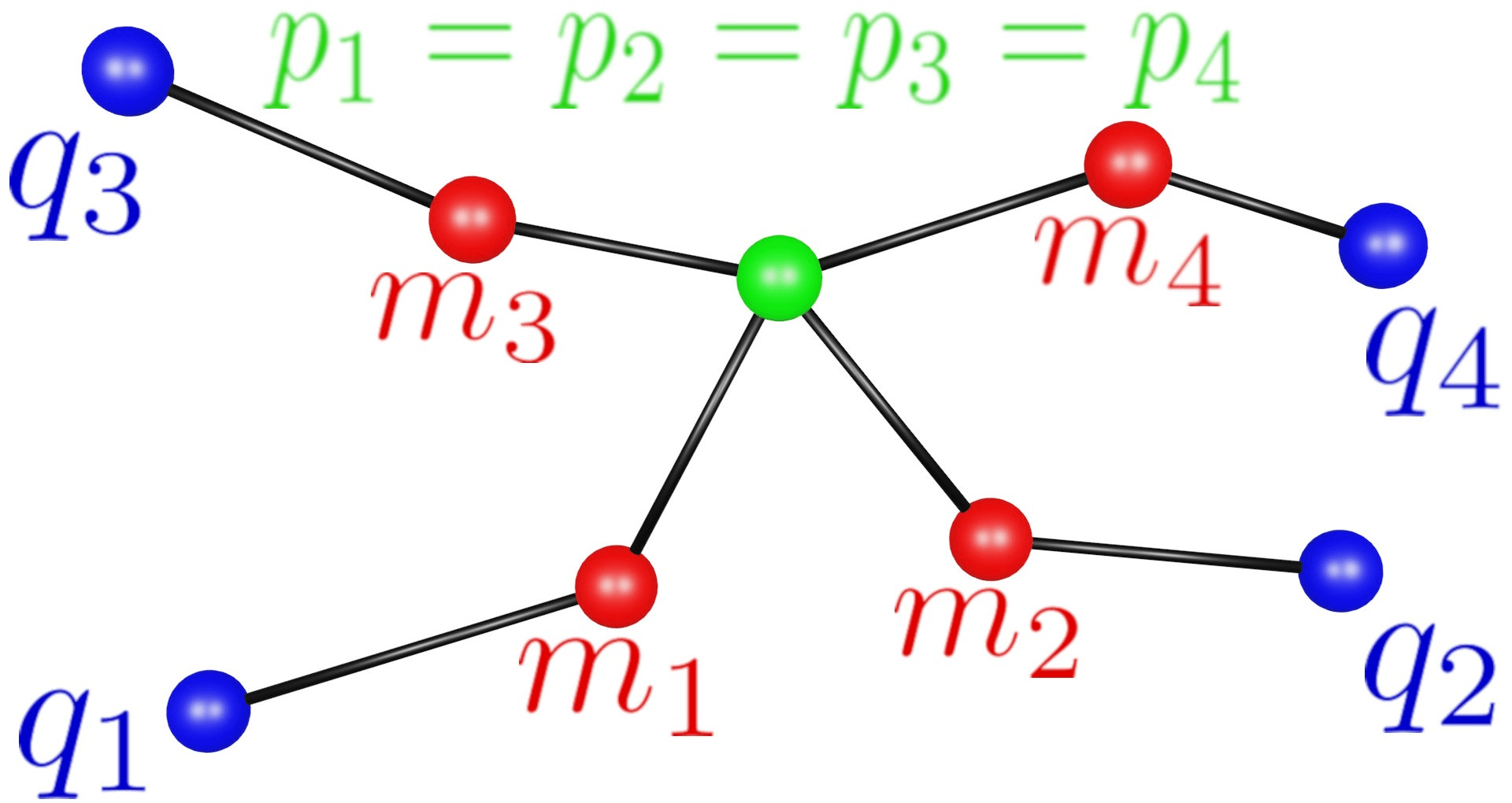}
    \end{subfigure}
    \hfill
    \caption{1-crank, 2-crank, and 4-crank. We have three types of points: the fixed ones ($q_i$ in blue), the joint points ($m_i$ in red), and the ``free'' point ($p_i$ in green).}
    \label{fig:configurations1}
\end{figure}

Let us apply the previous technique to the so-called linkages, allowing us to build more intricate C\=/spaces. Linkages are links (rigid bars) joined by joints (with ideal rotation) that can be placed at the end of the links or in the middle. The links can have some fixed points at the end. For instance, a double pendulum is a linkage with two links, one joint joining them, one fixed point at the beginning of one of the links and one free point at the end of the other. A pair of scissors is, simplifying, formed by two links, one joint placed in the middle of both links, and four free ends. Linkages can be further classified and are essential in engineering, but it is enough to focus on $n$-cranks for our purposes. We define a $1$-crank $C(q_1,L_1,\ell_1)$ as a double planar pendulum with lengths $L_1>\ell_1$ and whose fixed point is $q_1$. We denote its free end by $p_1$. A $2$-crank is formed by two $1$-cranks $\{C(q_i,L_i,\ell_i)\}_{i=1,2}$ joined by the free end i.e. $p_1=p_2$. A $3$-crank is given by three $1$-cranks where all the free ends $p_i$ are merged in a joint. Analogously, we can define the $n$-crank (see Fig.~\ref{fig:configurations1}).

We start by showing that the number of DOF of an $n$-crank is $2$. First, each crank has a middle point $m_i$, adding two DOFs but also two constraints (the distance from the fixed point $q_i$ to the middle point $m_i$ and the distance from $m_i$ to the endpoint $p_i$). Hence, it does not contribute to the DOFs. Second, the endpoint $p_i$ adds two DOFs, and since all $p_i$ are identified, two is the total number of DOFs.

\subsubsection{2-crank}\label{subsubsection 2 crank}

To study the $2$-crank, let us briefly recall what we did in the previous section for the double pendulum ($1$-crank). First, we derived that the R-space of $p_1$ is the space between the max-circle and the min-circle around the fixed point $q_1$. We denote from now on that closed region as $\mathscr{A}(q_1,L_1,\ell_1)$, the interior as $A(q_1,L_1,\ell_1)$, and its boundaries as $\S^1(q_1,r_1^\pm)$ where $r_1^\pm:=L_1\pm \ell_1$. On the one hand, we have one configuration per boundary point: $\straight$ for the max-circle and $\antistraight$ for the min-circle. On the other hand, we have two configurations per interior point: $\clock$ and $\counterclock$. This leads to a C-space formed by two annuli where the boundaries are identified, leading to a torus as depicted in Fig.~\ref{fig:annulusToTorus}. 

\begin{figure}[!ht]
\centering
     \begin{subfigure}[b]{0.45\textwidth}
         \centering
         \includegraphics[height=.45\textwidth]{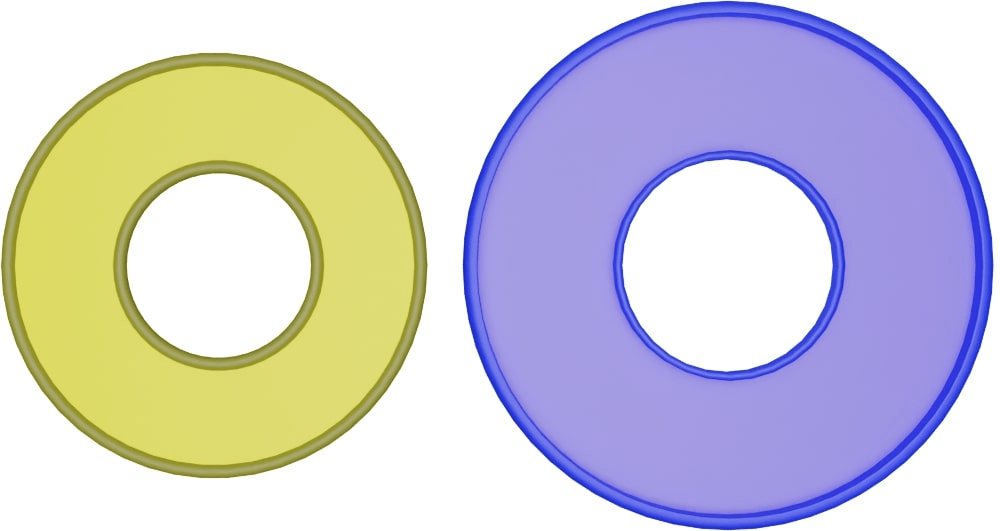}
         \caption{Too far apart. Empty intersection.} \label{empty intersection}
     \end{subfigure}
     \begin{subfigure}[b]{0.45\textwidth}
         \centering
         \includegraphics[height=.45\textwidth]{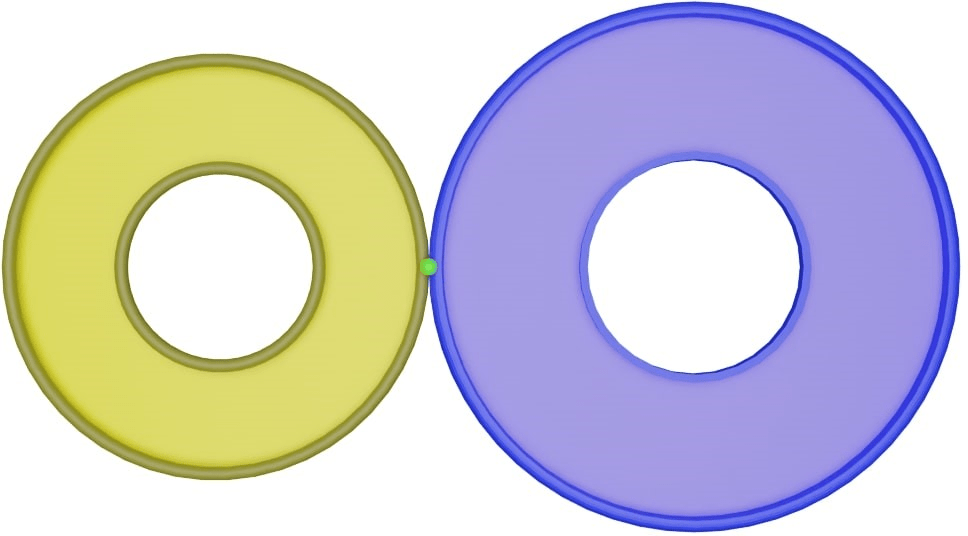}
         \caption{Only one possible configuration.} \label{one point intersection}
     \end{subfigure}\\
     \begin{subfigure}[b]{0.45\textwidth}\vspace{2ex}
         \centering
         \includegraphics[height=.45\textwidth]{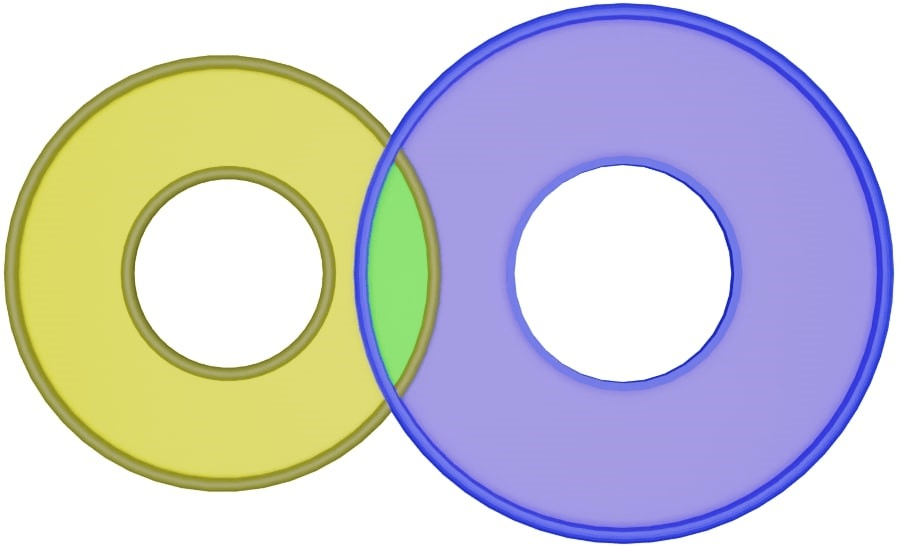}
         \caption{Oval or $2$-agon (two sides).}\label{oval}
     \end{subfigure}%
     \begin{subfigure}[b]{0.45\textwidth}
         \centering
         \includegraphics[height=.45\textwidth]{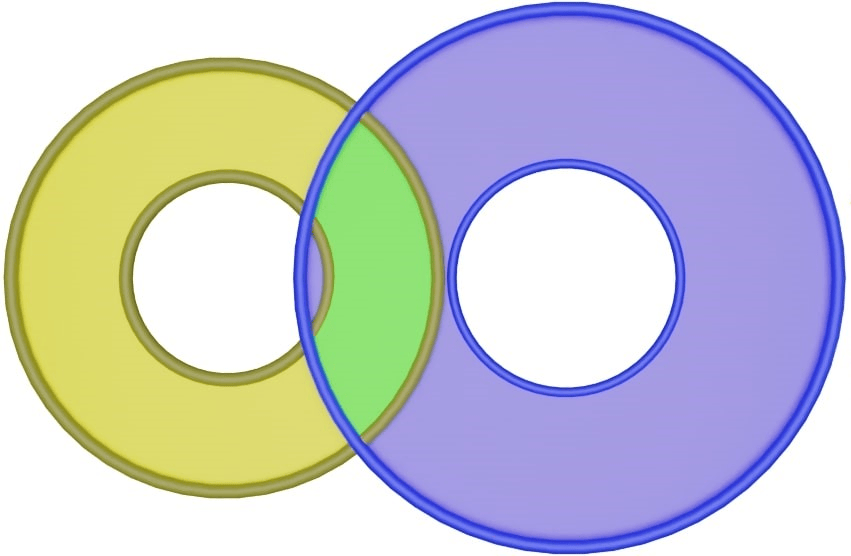}
         \caption{$4$-agon (four sides).} \label{Fig: 4-agon}
     \end{subfigure}%
     \\
     \begin{subfigure}[b]{0.45\textwidth}\vspace{2.5ex}
         \centering
         \includegraphics[height=.45\textwidth]{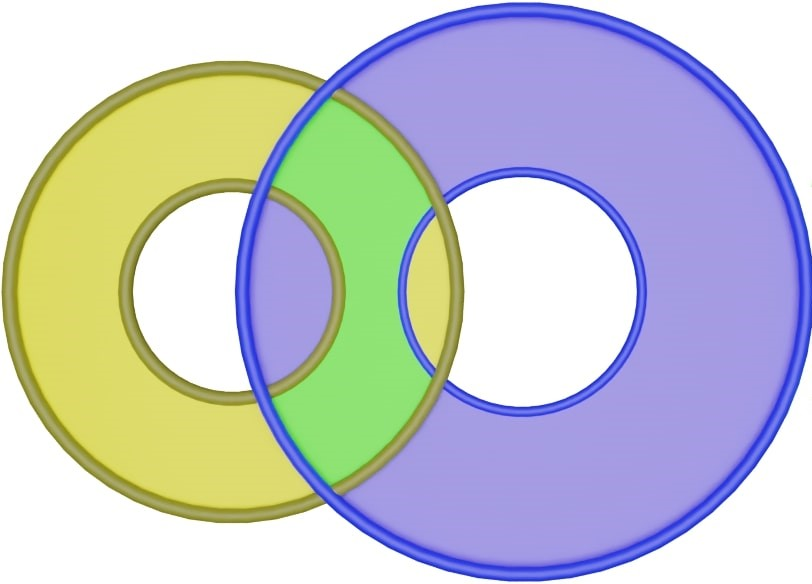}
         \caption{$6$-agon (six sides).}  \label{Fig:6-agon}
     \end{subfigure}%
     \begin{subfigure}[b]{0.45\textwidth}
         \centering
         \includegraphics[height=.45\textwidth]{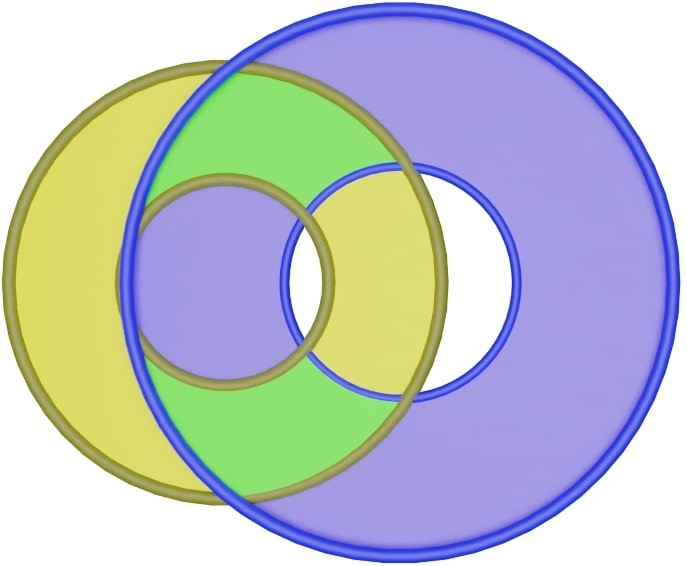}
         \caption{More intricate R-space.} \label{Fig:disjoint}
     \end{subfigure}
    
    \caption{The left annulus (yellow) is the R-space of the first pendulum, and the right annulus (blue) is the R-space of the second pendulum. The intersection, highlighted in each image (green), represents $\mathscr{R}_2$, the R-space of the 2-crank.}
    \label{intersection 2-crank}
\end{figure}

For the $2$-crank, we can repeat the process twice and look for the intersection. Indeed, $p_i$ can reach every point in $\mathscr{A}(q_i,L_i,\ell_i)$ but since $p_1=p_2$, the R-space of the $2$-crank is
\begin{equation}
\mathscr{R}_2:=\mathscr{A}\!\left(q_1,L_1,\ell_1\right)\cap \mathscr{A}\!\left(q_2,L_2,\ell_2\right)
\end{equation}
which can take different shapes depending on the parameters involved $(q_i,L_i,\ell_i)$. In particular, it could be empty if $q_1$ is too far apart from $q_2$ (Fig.~\ref{empty intersection}). It could be just a point if the distance between the centers is precisely $L_1+\ell_1+L_2+\ell_2$ (Fig.~\ref{one point intersection}). Or it could have several connected components (see the intersection region of Fig.~\ref{Fig:disjoint} in green).

Let us study the possible configurations for each point $p$ of $\mathscr{R}_2$. In the interior, each crank can be in the $\clock$ or $\counterclock$ configuration. If $p$ lies in the boundary of $\mathscr{R}_2$ but is not a corner, then one crank has two possible configurations while the other is locked in one configuration ($\straight$~for the max-circle and $\antistraight$ for the min-circle, see Fig.~\ref{fig:configurations}). If $p$ is one of the corners of $\mathscr{R}_2$, then both cranks are locked, and only one configuration is available. Overall, we have $4\times 4=16$ possible configurations summarized in Table~\ref{table:ConfigurationTable}.
\begin{longtable}{|c|c|wl{6.3cm}|}\hline
\label{table:ConfigurationTable}
Possible configurations & \multirow{2}{*}{Region} & \multirow{2}{*}{\qquad\quad Further explanation}\\
$($1st crank, 2nd crank$)$&&\\
\hline\hline\rule{0ex}{3ex}
\multirow{3}{*}{\shortstack{$(\clock,\clock)$\qquad $(\clock,\counterclock)$\\ $(\counterclock,\clock)$\qquad $(\counterclock,\counterclock)$}} & & Both crank endpoints located in the\hfill\mbox{}\\
 & $A\!\left(q_1,L_1,\ell_1\right)\cap A\!\left(q_2,L_2,\ell_2\right)$ & interior of their respective annulus\hfill\mbox{}\\
 &  & (each crank has 2 possible configurations)\hfill\mbox{}
\\[1ex]\hline\rule{0ex}{3ex}
$(\clock,\straight)$ &   \multirow{2}{*}{$A\!\left(q_1,L_1,\ell_1\right)\cap     \S^1(q_2,r_2^+)$}& 1st crank in the interior (2 conf.)\\[.5ex]
$(\counterclock,\straight)$&& 2nd one over max-circle (1 conf.)\\[1ex]\hline\rule{0ex}{3ex}
$(\clock,\antistraight)$ &     \multirow{2}{*}{$A\!\left(q_1,L_1,\ell_1\right)\cap     \S^1(q_2,r_2^-)$}& 1st crank in the interior (2 conf.)\\[.5ex]
$(\counterclock,\antistraight)$&&2nd one over min-circle (1 conf.)\\[1ex]\hline\rule{0ex}{3ex}
$(\straight,\clock)$ &   \multirow{2}{*}{$\S^1(q_1,r_1^+)\cap A\!\left(q_2,L_2,\ell_2\right)    $}& 1st crank over max-circle (1 conf.)\\[.5ex]
$(\straight,\counterclock)$&& 2nd one in the interior (2 conf.)\\[1ex]\hline\rule{0ex}{3ex}
$(\antistraight,\clock)$ &     \multirow{2}{*}{$\S^1(q_1,r_1^-)A\cap \!\left(q_2,L_2,\ell_2\right)    $}& 1st crank over min-circle (1 conf.)\\[.5ex]
$(\antistraight,\counterclock)$&&2nd one in the interior (2 conf.)\\[1ex]\hline\rule{0ex}{3ex}
$(\straight,\straight)$        &    $\S^1(q_1,r_1^+)\cap     \S^1(q_2,r_2^+)$& Both max-circles (1 conf. each).\\[1ex]\hline\rule{0ex}{3ex}
$(\straight,\antistraight)$    &     $\S^1(q_1,r_1^+)\cap     \S^1(q_2,r_2^-)$&Max-circle and min-circle (1 conf. each)\\[1ex]\hline\rule{0ex}{3ex}
$(\antistraight,\straight)$      & $\S^1(q_1,r_1^-)  \cap   \S^1(q_2,r_2^+)$& Min-circle and max-circle (1 conf. each)\\[1ex]\hline\rule{0ex}{3ex}
$(\antistraight,\antistraight)$  &    $\S^1(q_1,r_1^-)  \cap   \S^1(q_2,r_2^-)$& Both min-circles (1 conf. each)\\[1ex]\hline 
\caption{Possible configurations that appear when the free point moves around the R-space of the $2$-crank. See Fig.~\ref{intersection 2-crank} for some possible R-spaces.}
\end{longtable}
Some of the regions of this table can be empty or just to a point. Let us explore a couple of examples and determine the C-space. 

\begin{figure}[!ht]
\centering\includegraphics[width=0.71\textwidth]{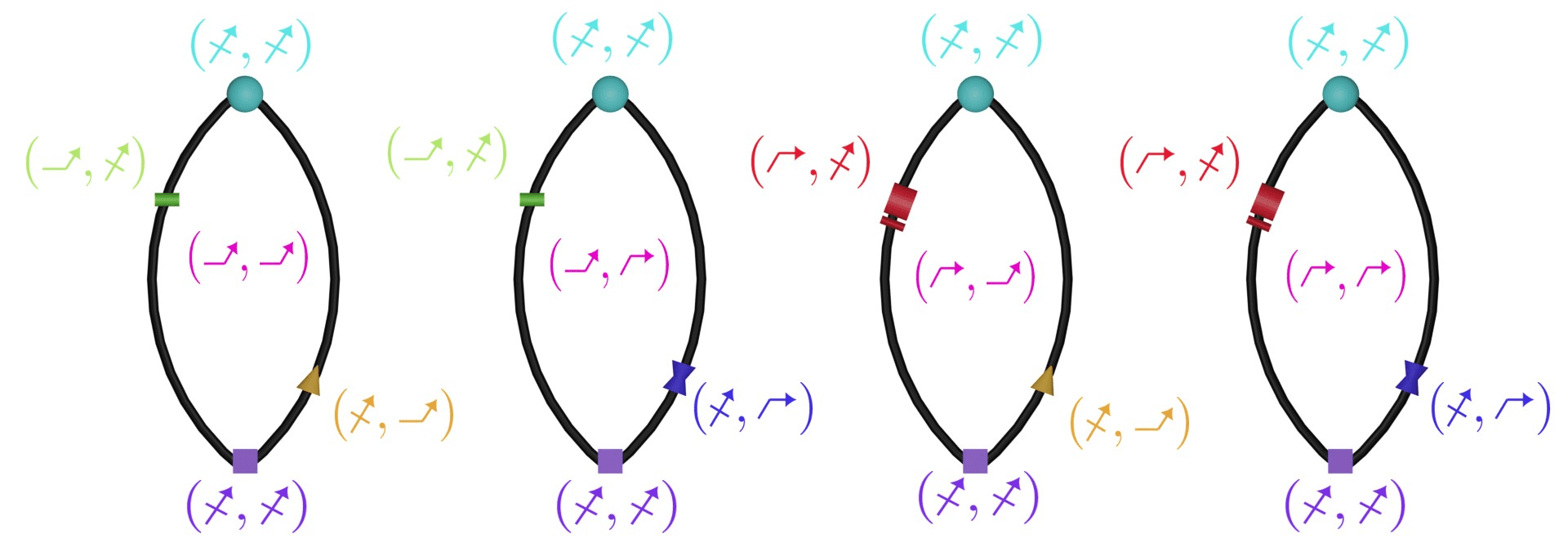}
\caption{Four ovals corresponding to the possible configurations in the interior.}\label{Fig. four ovals}
\end{figure}

\textbf{2-crank - Gluing for the $2$-agon case}\\
We already mentioned that if $\mathrm{d}(q_1,q_2)=r^+_1+r^+_2$, the only possible configuration is $(\straight,\straight)$ and the C-space is just a point (Fig.~\ref{one point intersection}). The next case to consider is when $\mathrm{d}(q_1,q_2)$ is slightly smaller, as depicted in Fig.~\ref{oval} where the R-space forms an oval (or $2$-agon since it has two sides). Mathematically, this condition reads:
\begin{equation}\label{eq: distance 2-crank 1}
\mathrm{max}\{r^+_1+r^-_2,r^-_1+r^+_1\}<\mathrm{d}(q_1,q_2)<r^+_1+r^+_2 \; . 
\end{equation}
The second inequality ensures that both cranks can be bent, while the first one ensures that none of the min-circles are reachable, i.e., the $\antistraight$ configuration is not available to either of the cranks. Removing such configurations from the table leaves us with a C-space formed by four ovals whose edges have at least one $\straight$ configuration, as Fig.~\ref{Fig. four ovals} shows.

\begin{figure}[!ht]
\centering\includegraphics[width=.24\textwidth]{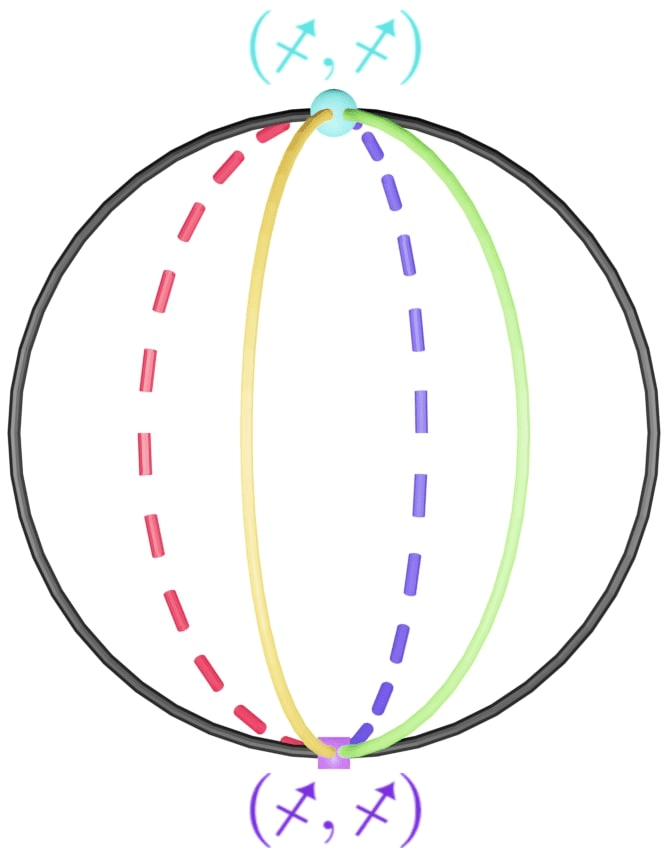}\caption{Four ovals glued together to form a sphere. Notice that the north and south poles correspond to the vertices of the ovals from Fig.~\ref{Fig. four ovals}.}\label{Fig. sphere from ovals}
\end{figure}

To understand better the resulting C-space, we proceed as we did for Fig.~\ref{fig:annulusToTorus}. We glue the equivalent edges $(\clock,\straight)$ of the first two ovals (notice that one has to be flipped) and analogously for the edges $(\counterclock,\straight)$ of the last two ovals. This leads to two disks with equivalent boundaries, and by gluing them, we obtain that the C-space is a sphere (see Fig. \ref{Fig. sphere from ovals}).
 
\textbf{2-crank - Gluing for the $4$-agon case}\\
The next case to consider is when the centers are separated such that the boundary of the R-space is formed by one of the min-circles and by both of the max-circles (see Fig.~\ref{Fig: 4-agon}):
\begin{equation}\label{eq: distance 2-crank 2}
r^-_1<\mathrm{d}(q_1,q_2)-r^+_2<\mathrm{d}(q_1,q_2)-r^-_2<r^+_1<\mathrm{d}(q_1,q_2)
\end{equation}
Then, we end up with four $4$-agons. 

\begin{figure}[!ht]
\centering\includegraphics[width=0.85\textwidth]{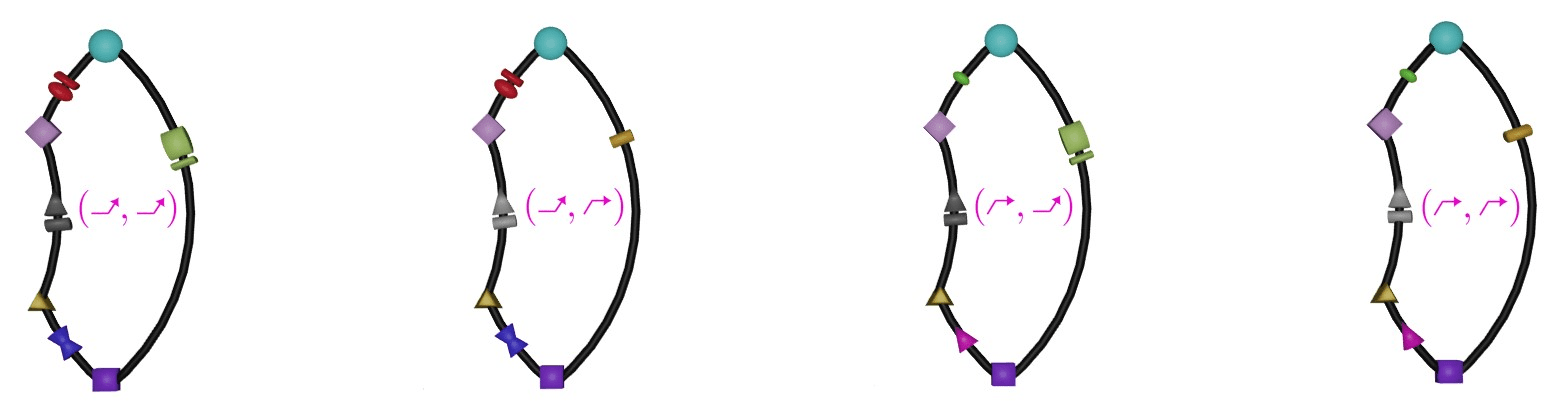}
\caption{Four $4$-agons which correspond to the $2\times 2$ possible configurations in the interior. Notice that they can be thought of as the previous four ovals from Fig. \ref{Fig. four ovals} with a piece of disk removed (see also Fig. \ref{Fig: 4-agon}).}\label{four 4-agons}
\end{figure}

Gluing as before, leads to two annuli with the boundaries identified. Thus, the C-space is a torus.

The avid reader might have realized that the C-space has changed its topology from a sphere to a torus by continuously decreasing $\mathrm{d}(q_1,q_2)$. This change can be understood as a ``phase transition'' where the inner hole of the torus collapses to a point when the cranks get closer or, conversely, where we create a puncture on the sphere when the cranks get further apart.

\textbf{2-crank - Gluing for the $6$-agon case}\\
We leave as an exercise to prove, using the same techniques, that the C-space of the case where R-space is a $6$-agon (see Fig.~\ref{Fig:6-agon}), which mathematically reads
\begin{equation}
    0<\mathrm{d}(q_1,q_2)-r^+_2<r^-_1<\mathrm{d}(q_1,q_2)-r^-_2<r^+_1<\mathrm{d}(q_1,q_2)\, ,
\end{equation}
is a topological surface with two holes. Notice that a $4$-agon can be thought, as explained in the caption of Fig.~\ref{four 4-agons}, as a $2$-agon with a piece of a disk removed. That piece removed led, after gluing, to a hole in the sphere obtained for the $2$-agon. Thus, it makes sense that the $6$-agon, which can be thought of as a $4$-agon with another piece of disk removed, leads to a surface with two holes.

Although this gluing strategy to determine the C-space is very visual and easy to implement in these simple examples, it becomes harder when more cranks are involved. In that case, the calculation becomes much easier if we make use of some tools from topology. The reader familiar with the Euler characteristic of a surface is referred to the appendix, where these tools are introduced and applied to the 3-crank and $n$-crank.

\section{Conclusions}

DOFs are one of the key topics in a first course in classical mechanics. They are also essential in other areas, such as statistical mechanics, condensed matter, field theories and robotics. In this paper, we have focused on the DOFs of classical mechanics and show how they are more subtle than one might expect, and require some reflection to obtain a deeper knowledge and understanding. In particular, we have seen that we have to understand DOFs as coordinates on the configuration space. This means that one should be careful when assigning physical meaning to DOFs (especially when dealing with configuration spaces which are not manifolds) in the same way that one should be careful when assigning meaning to coordinates.


This perspective of understanding degrees of freedom as coordinates of the configuration space has an additional advantage: it can be used to build a better intuition for manifolds and topology. Indeed, some surfaces with non-trivial topologies can be realized as mechanical systems as well as some higher dimensional manifolds. In fact, this approach provides an excellent motivation for a topology or differential geometry course for physicists.

\section*{Acknowledgements}

JMB was supported by a fellowship from the Atlantic Association of the Mathematical Sciences (AARMS), the Natural Sciences and Engineering Research Council of Canada (NSERC) Discovery Grants 2018-04887 and 2018-04873, and by the Spanish Ministerio de Ciencia Innovación y Universidades and the Agencia Estatal de Investigación PDI2020-116567GB-C22. IB was also supported by NSERC grant 2018-04873. The lands on which Memorial University’s campuses are situated are in the traditional territories of diverse Indigenous groups, and we acknowledge with respect the diverse histories and cultures of the Beothuk, Mi’kmaq, Innu, and Inuit of this province.

\begin{appendices}

\mbox{}

\section{C-spaces and the Euler characteristic}

\subsection{Euler characteristic and classification of surfaces}
In this appendix, we present an alternative derivation to obtain the topology of the C-space of an \mbox{$n$-crank} that relies on the fact that the topology of compact surfaces is completely characterized by two parameters: the orientability and the genus of the surface. The proof of this fact can be found in any algebraic topology or surface topology book like \cite[Ch.12]{munkres2000topology,nakahara2003geometry}. The genus, in turn, is characterized by the Euler characteristic $\rchi$, which can be computed following these steps for any given surface:

\begin{enumerate}
    \item Decompose the surface into polygons.
    \item Count the number of polygons (faces) $F$, the number of edges $E$, and the number of vertices $V$.
    \item Compute its alternate sum $\rchi=F-E+V$
\end{enumerate}

It can be proved that this number does not depend on the decomposition (e.g., if we break a triangle in two, we increase $F$ by one, $E$ by 2, and $V$ by 1, so $\rchi$ does not change).  Moreover, we have
\begin{equation}
    \rchi=\begin{cases}
     2(1-g)& \text{If the surface is orientable} \\
     2-g& \text{If the surface is non-orientable}
\end{cases}
\end{equation}
where $g$ is the genus of the surface. For orientable surfaces, the $g$ is the number of holes: zero for the sphere, 1 for the torus, and so on. For non-orientable surfaces, it is a bit more complicated. Luckily, all the surfaces in this section are orientable (though see Appendix \ref{appendix more cspaces} for a non-orientable example), and so we have the concise formula 
\begin{equation}\label{eq:genus}
    g=\frac{1}{2}(2-F+E-V) \; .
\end{equation}

\begin{figure}[ht]
     \centering
     \begin{subfigure}[b]{0.11\textwidth}
         \includegraphics[width=\textwidth]{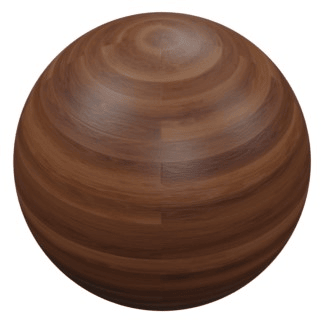}
         \caption{g = 0}
         \label{fig:Sphere-GenusZero}
     \end{subfigure}
     \hfill
     \begin{subfigure}[b]{0.16\textwidth}
         \includegraphics[width=\textwidth,height=5em]{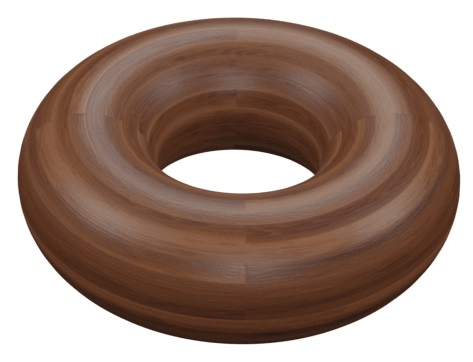}
         \caption{g = 1}
         \label{fig:1-torus}
     \end{subfigure}
     \hfill
     \begin{subfigure}[b]{0.22\textwidth}
         \includegraphics[width=\textwidth,height=5em]{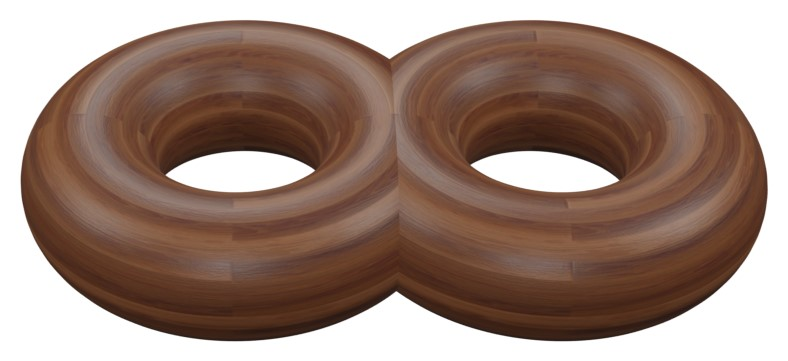}
         \caption{g = 2}
         \label{fig:2-torus}
     \end{subfigure}
     \hfill
     \begin{subfigure}[b]{0.22\textwidth}
         \includegraphics[width=\textwidth,height=5em]{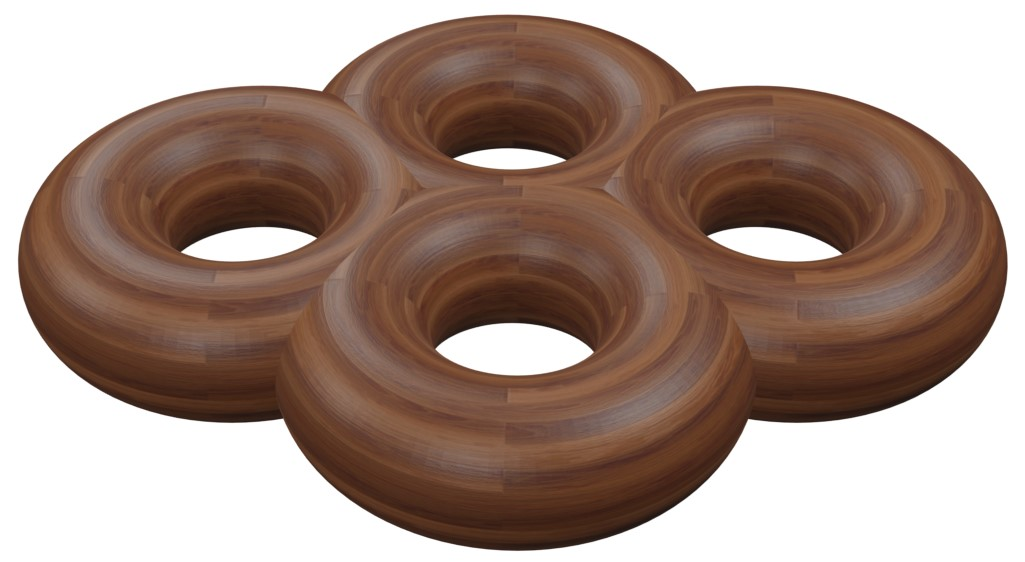}
         \caption{g = 4}
         \label{fig:4-torus}
     \end{subfigure}
     \caption{Several topological orientable manifolds with different genera.}\label{fig:genera}
\end{figure}

\subsection{2-crank revisited}

Let us apply equation \eqref{eq:genus} to the $2$-crank. In the $2$-agon case (Fig.~\ref{Fig. four ovals} and Fig.~\ref{Fig. sphere from ovals}), we have $F=1\times 4$ (one for each oval), $E=2\times 4/2=4$ (there are two sides for each of the 4 ovals, but half of them are the same since they are identified), and $V=2\times 4/4=2$ (the top vertex of every oval is identified and the same for the bottom one), so the genus is $g=0$ and this implies that the surface is a sphere.

For the $4$-agon case (see Fig.~\ref{four 4-agons}), a similar count leads to $(F,E,V)=(4,8,4)$, so $\rchi=0$. As we have mentioned before, it can be proved that surfaces obtained through this gluing process for $n$-cranks are orientable. Hence, $g=1$, and the surface is a torus. Finally, the $6$-agon case we left as an exercise for the reader can be quickly solved. A quick count leads to $(F,E,V)=(4,12,6)$. Thus $g=2$. 

\textbf{2-crank - General case}\\
Let us deduce some general properties. Firstly, we can study each connected component separately (see, for instance, Fig.~\ref{Fig:disjoint}), since they define disconnected manifolds. Thus, the number of faces $F$ is $1$ times the number of distinct configurations in the interior of the R-space. Here, we have two possible configurations for each crank, so $F=2^2$. Secondly, the number of edges of the R-space, that we denote $k$, is variable, but over each edge, one of the cranks is locked in the $\straight$ or $\antistraight$ configuration, while the other crank has two configurations. Hence $E=2^{2-1}k$. Finally, the number of vertices of the R-space is the same as the number of edges (it is a loop), but both cranks are fixed into the $\straight$ or $\antistraight$ configuration, hence $V= 2^{2-2}k$. The C-space of a $2$-crank is an orientable surface of genus
\begin{equation}
    g=\frac{k}{2}-1\label{eq_genus2crank}
\end{equation}
for some $k$ to be found in each example. Notice that this equation is only valid in the generic case but not for the in-between cases (e.g., considering the equality in \eqref{eq: distance 2-crank 2} leads to an odd $k$). Those correspond to the ``phase transition'' mentioned earlier where the number of holes changes. We will see an example in the next section.

\subsection{3-crank}

Let us focus our attention on the $3$-crank. It is interesting to note that one particular configuration of this system (see Fig.~\ref{3crank3} below) was studied by Thurston and Weeks in \cite{thurston1984mathematics} and was a big inspiration to write this paper. We highly recommend that paper to the interested reader. The R-space of a $3$-crank is given by
\begin{equation}
    \mathscr{R}_3:=\mathscr{A}\!\left(q_1,L_1,\ell_1\right)\cap \mathscr{A}\!\left(q_2,L_2,\ell_2\right)\cap \mathscr{A}\!\left(q_3,L_3,\ell_3\right)
\end{equation}
Although we are not going to build a detailed table of all possible configurations like before, since it would be far too lengthy for this discussion, notice that the $4\times 4\times 4=64$ possible configurations are 
\begin{equation}
\{(x_1,x_2,x_3)\text{ where }x_i\in\{\clock,\counterclock,\straight,\antistraight\}\}
\end{equation}
Some of the possible regions are shown in Fig.~\ref{intersection 3-crank}.

\begin{figure}[!ht]
    \centering
     \begin{subfigure}[b]{0.5\textwidth}
         \centering
         \includegraphics[width=.56\textwidth]{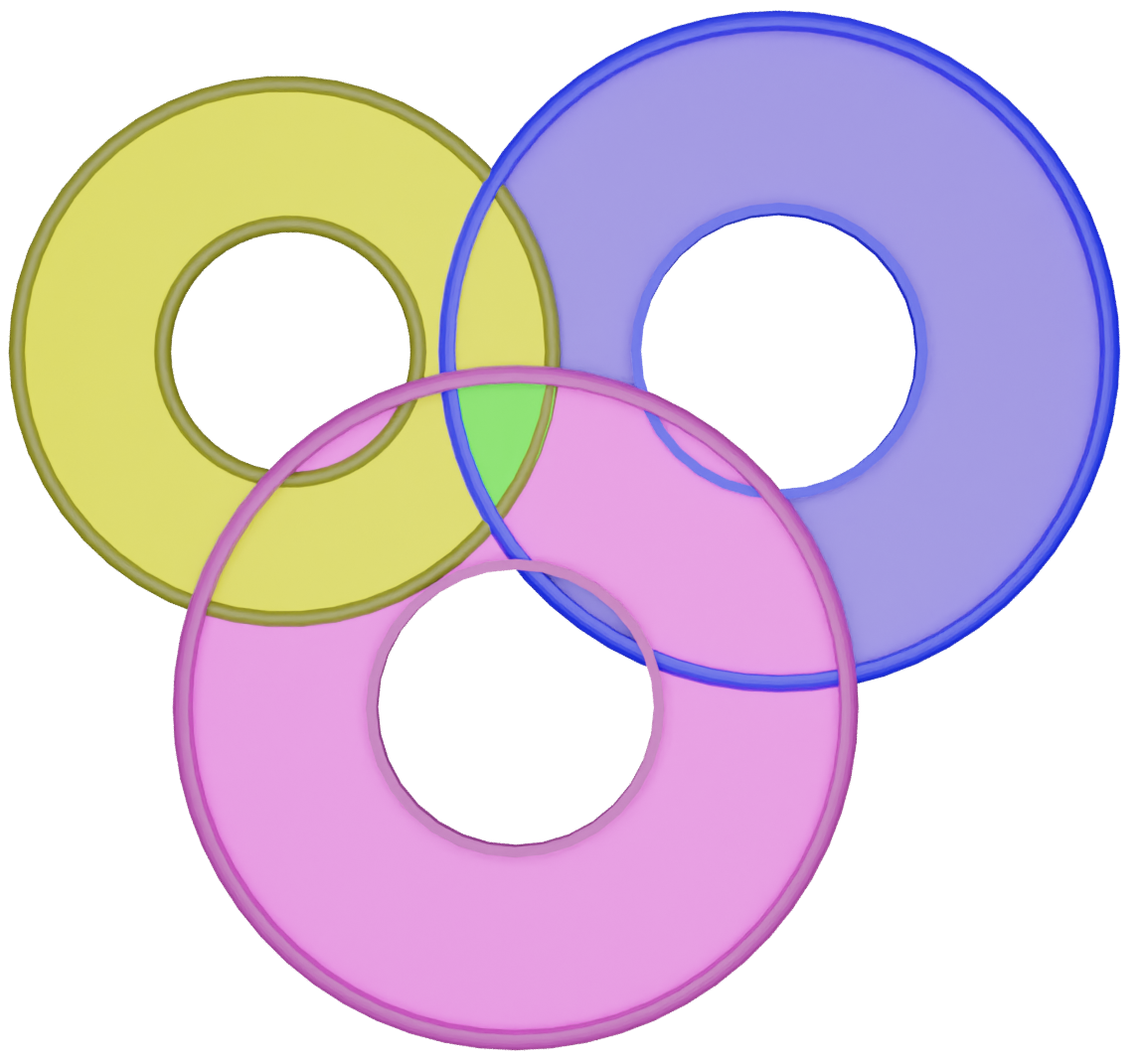}
         \caption{$3$-agon, only max-circles involved.\rule{-11ex}{-12ex}} \label{3crank1}
     \end{subfigure}%
     \begin{subfigure}[b]{0.5\textwidth}
         \centering
         \includegraphics[width=.56\textwidth]{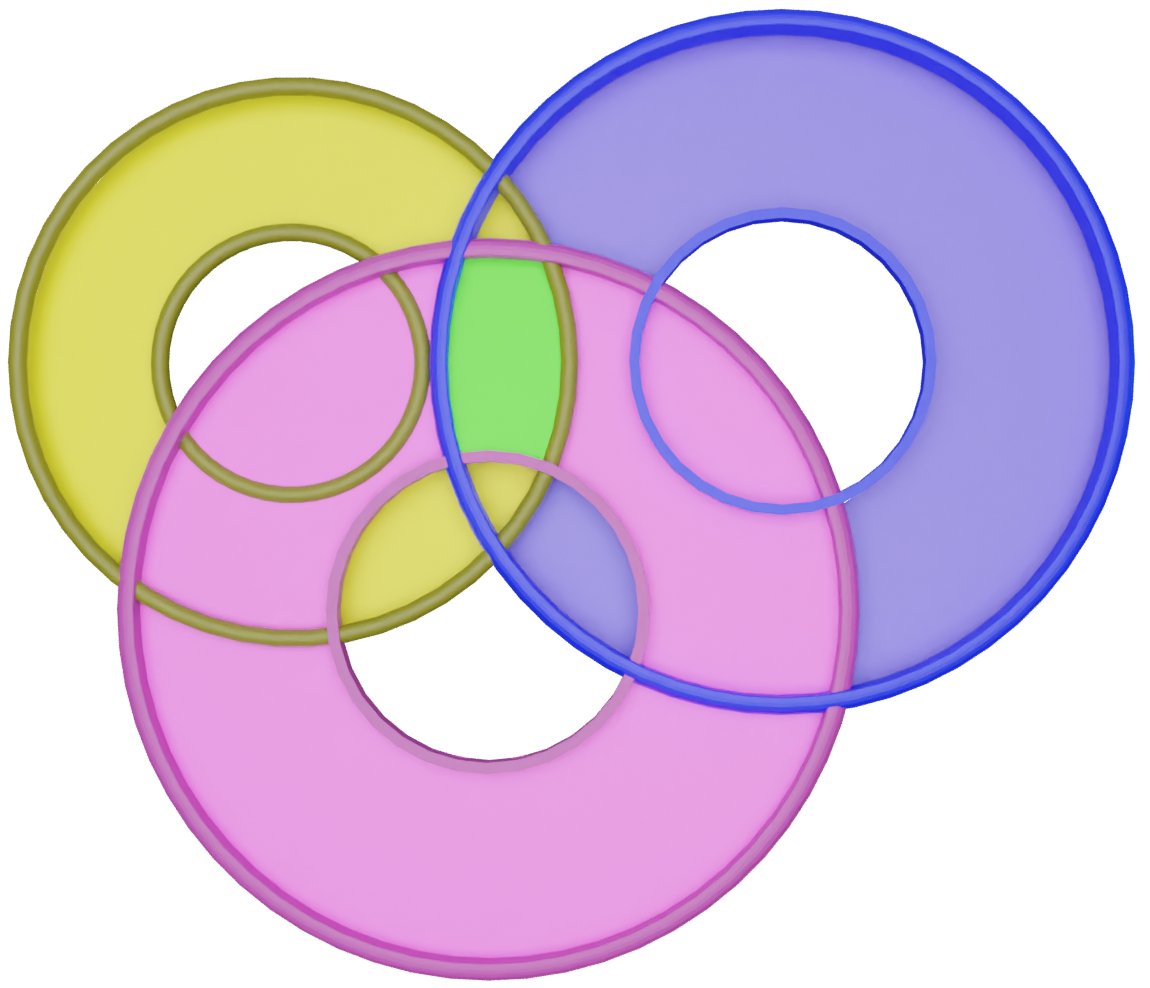}
         \caption{$4$-agon, only one min-circle involved.} \label{3crank2}
     \end{subfigure}%
\end{figure}%
\begin{figure}[ht]\ContinuedFloat
     \begin{subfigure}[b]{0.5\textwidth}
         \centering
         \includegraphics[width=.56\textwidth]{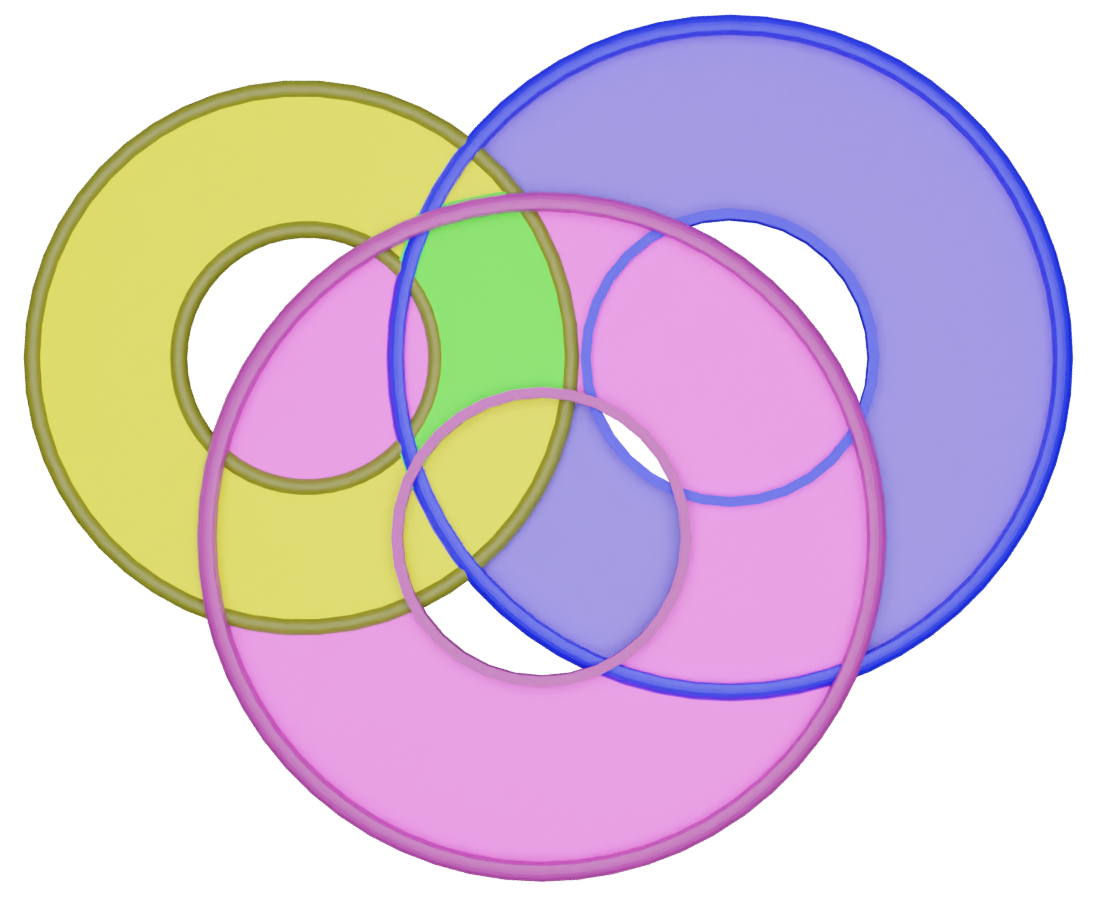}
         \caption{$6$-agon, every min-circle and max-circle involved.}\label{3crank3}
     \end{subfigure}%
     \begin{subfigure}[b]{0.5\textwidth}
         \centering
         \includegraphics[width=.56\textwidth]{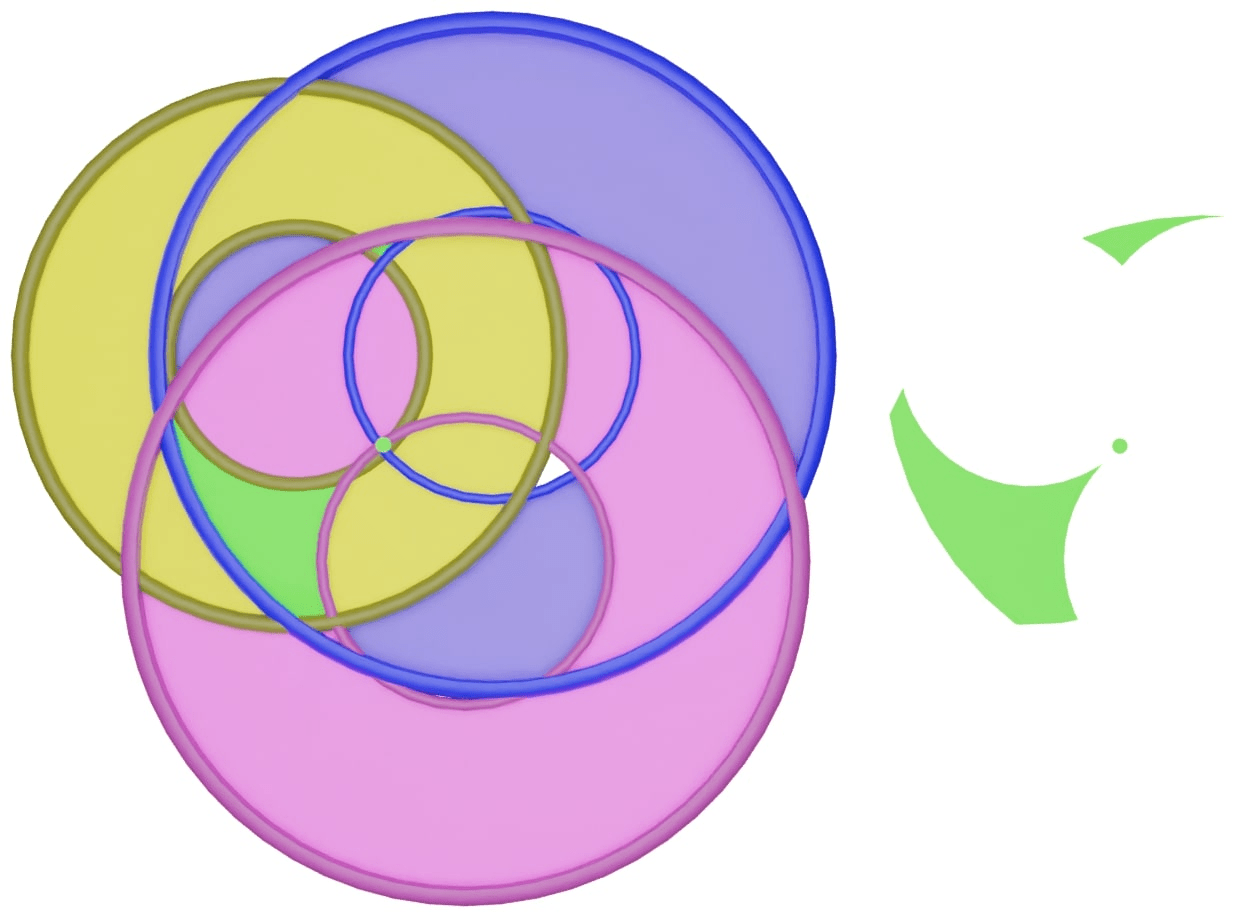}
         \caption{A more intricate R-space.} \label{3crank4}
     \end{subfigure}%
        \caption{The top-left annulus (yellow) of each image represents the R-space of the first pendulum. The top-right annulus (blue), the R-space of the second pendulum. The bottom one (pink), the R-space of the third pendulum. The intersection is highlighted in each image (green). Moreover, in Fig. (d), we have reproduced the R-space on the right for easier visualization.} \label{intersection 3-crank}
\end{figure} 

\textbf{3-crank - Euler characteristic for the general (nondegenerate) case}\\
Each connected component of the R-space is a $k$-agon. Then,
\[\begin{array}{l}
F=1\times2^3=8\\
E=k\times 2^{3-1}=4k\\
V=k\times 2^{3-2}=2k
\end{array}\qquad\overset{\text{Eq.~\eqref{eq:genus}}}{\xrightarrow{\hspace*{2.5cm}}}\qquad g=k-3\]
In \cite{thurston1984mathematics}, the authors study the case $k=6$ and prove, through the gluing process, that the surface has 3 holes, which agrees with our quick computation.

\begin{figure}[!ht]
    \centering\includegraphics[clip,trim=0 0 110ex 0,width=.38\textwidth]{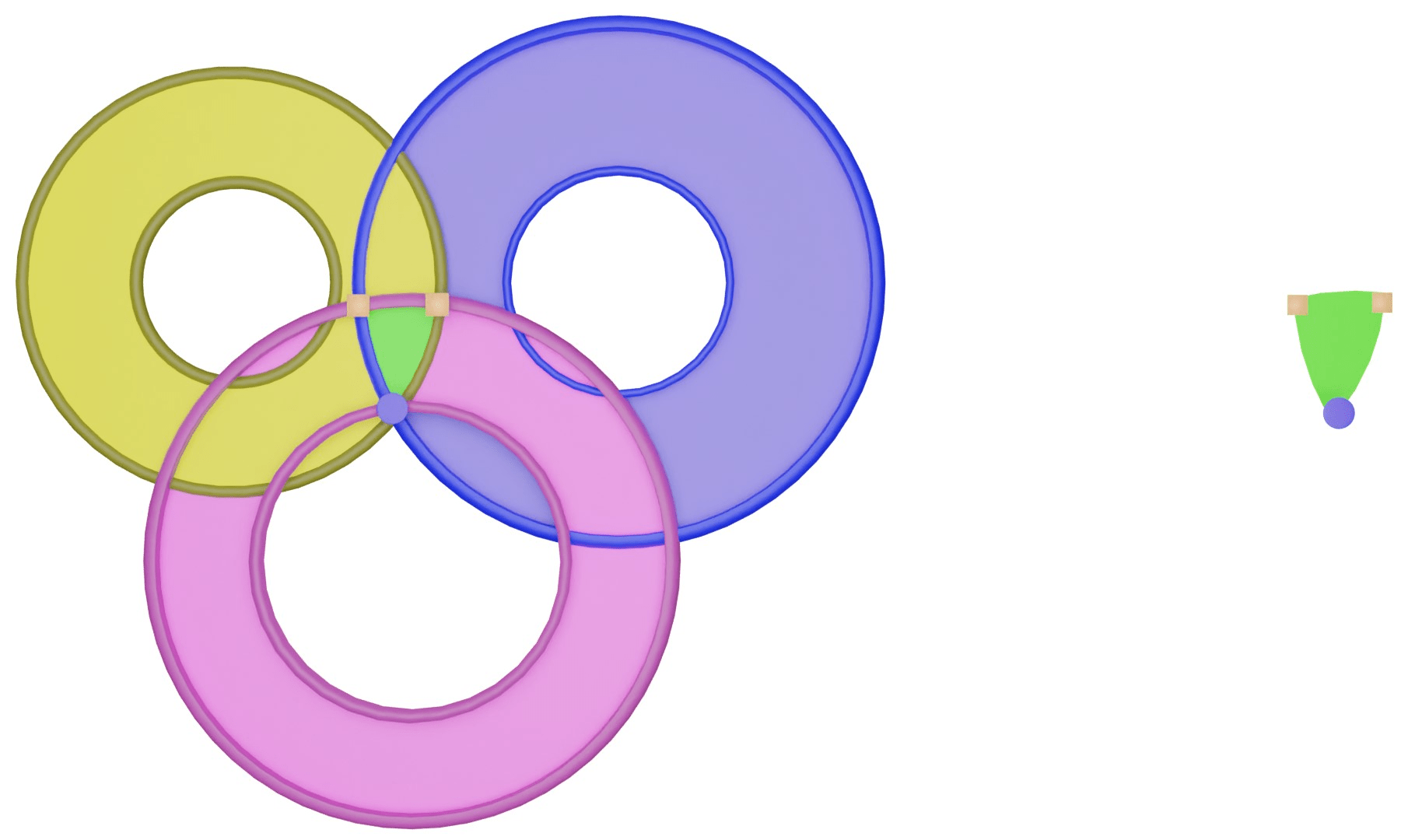}
    \caption{3-crank of a degenerate case where there is a point of the R-space where all three cranks are locked into the straight position.}\label{Fig. degenerate 3-crank}
\end{figure}

\textbf{3-crank - A degenerate case}\\
As we mentioned for the $2$-crank, the previous formula does not cover the degenerate cases, which must be dealt with separately. Consider, for instance, the configuration portrayed in Fig.~\ref{Fig. degenerate 3-crank}.

\begin{figure}[ht!]
\centering\includegraphics[width=.31\linewidth]{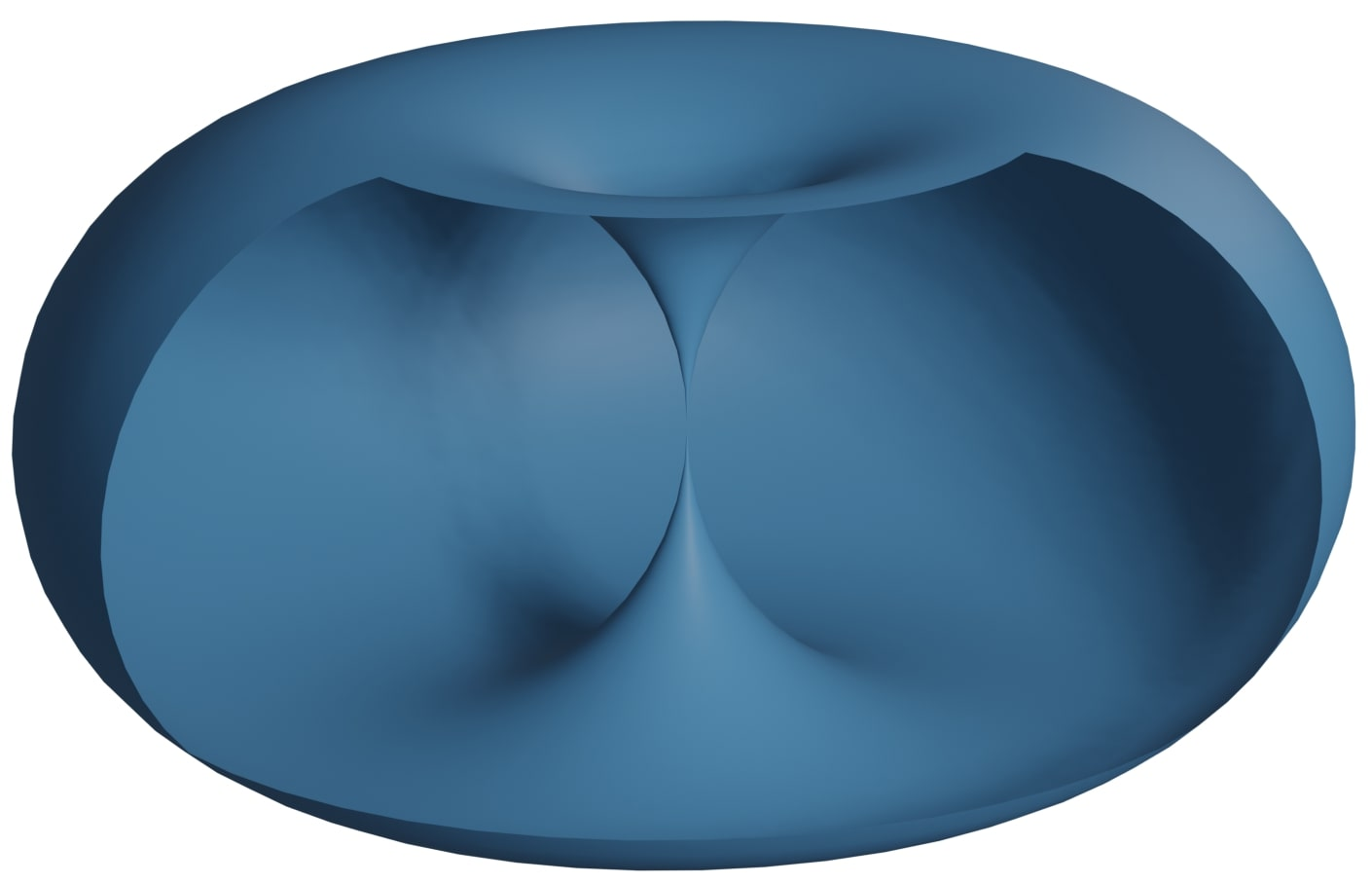}
    \caption{Horn torus obtained identifying the north and south poles of a sphere.}
    \label{fig:horn}
\end{figure}

This case might seem to be the same as the $3$-agon case, so $g=3-3=0$, i.e., a sphere. However, the bottom vertex (blue circle) differs from the top ones (orange squares). Indeed, the former is the intersection of three circles, meaning that all three cranks are locked, and the only possible configuration is $(\straight,\straight,\antistraight)$. The other vertices have one crank free, so they have two configurations each. Thus, the counting is
\[\begin{array}{l}
F=1\times2^3=8\\
E=3\times 2^{3-1}=12\\
V=2\times 2^{3-2}+1\times 2^{3-3}=5
\end{array}\qquad\overset{\text{Eq.~\eqref{eq:genus}}}{\xrightarrow{\hspace*{2.5cm}}}\qquad g=\frac{1}{2}\]
which is impossible as the genus is a natural number. Notice that this computation is telling us that, in a sense, the C-space is in between a sphere ($g=0$) and a torus ($g=1$). More specifically, we obtain a sphere with the north and south poles identified, leading to a horn torus (see Fig.~\ref{fig:horn}). A similar but more complicated situation occurs if two edges of a min-circle or max-circle are exactly the same.

\subsection{\texorpdfstring{$\textbf{n}$}{n}-crank}

In this section, we consider the $n$-crank whose R-space is
\begin{equation}
\mathscr{R}_n:=\bigcap_{i=1}^n\mathscr{A}\!\left(q_i,L_i,\ell_i\right)
\end{equation}
First, notice that if the max-circle and min-circle of the $i$-th crank are not part of the boundary of $\mathscr{R}_n$, then the $i$-th crank plays no role. We can compute the topology of the C-space ignoring that crank and then realizing that we have two disconnected copies of that C-space: one where the $i$-th crank is in the $\clock$ configuration and the other where it is in the $\counterclock$ configuration. Since $\S^1(q_i,r_\pm)\cap\mathscr{R}_n$ is empty, the $i$-th crank cannot change from one to the other without breaking the mechanical system.

\begin{figure}[h!]
    \centering
    \includegraphics[width=.435\textwidth]{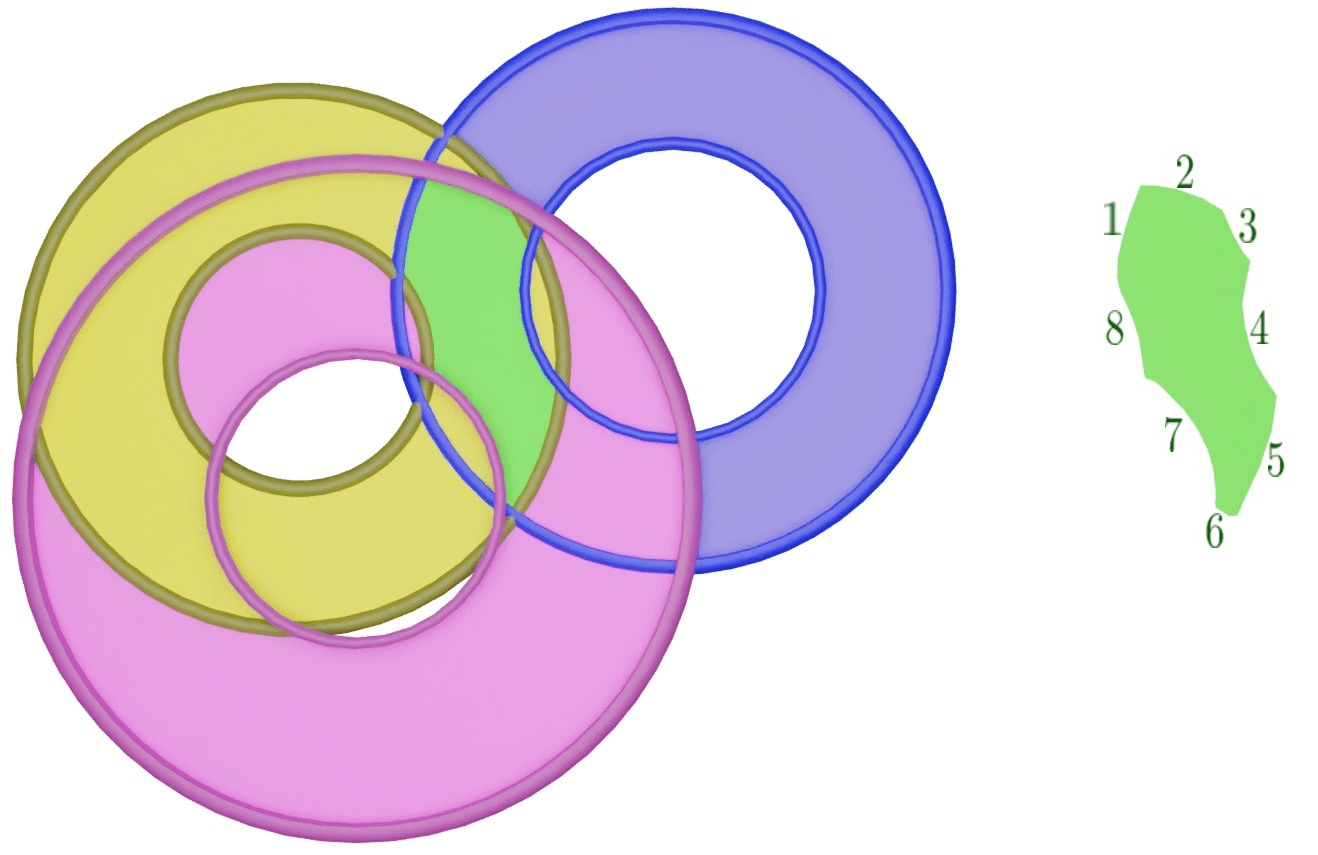}
    \caption{A 3-crank with an $8$-agon R-space (the intersection of the three annuli).}
    \label{fig:8-agon}
\end{figure}

Consider one of the connected components of $\mathscr{R}_n$, which we know is a $k$-agon. Since we are excluding the aforementioned degenerate cases, we know that each edge has $2^{n-1}$ configurations, and each vertex has $2^{n-2}$. Thus
\[\begin{array}{l}
F=2^n\\
E=2^{n-1}k\\
V=2^{n-2}k
\end{array}\qquad\overset{\text{Eq.~\eqref{eq:genus}}}{\xrightarrow{\hspace*{2.5cm}}}\qquad g=1+2^{n-3}(k-4)\]
Recall that we are assuming that at least one of the boundaries of each $\mathscr{A}\!\left(q_i,L_i,\ell_i\right)$ plays a role (otherwise, we can remove the $i$-th crank). Hence, $k$ is an integer larger than $n$. The upper bound $k_{\mathrm{max}}$ is not easy to derive. The naive guess of having at most $2n$ sides (every max-circle and every min-circle is involved once) is incorrect since we can have even more, as Fig.~\ref{fig:8-agon} suggests.

Here we have $k=8>2n=6$ and $g=5$. In any case, we have $k_{\mathrm{max}}\geq2n$, and we can say that the C-space is formed by connected components, each one of which is a surface with genus between
\begin{equation}
    g_{\mathrm{min}}=1+2^{n-3}(n-4)\qquad\text{ and }\qquad g_{\mathrm{max}}=1+2^{n-3}(k_{\mathrm{max}}-4)\geq1+2^{n-2}(n-2)
\end{equation}
as well as the degenerate intermediate cases (with one or more punctures as for the horn torus). Although we can realize a lot of surfaces as the C-space of an $n$-crank, not every possible surface can be realized. For instance, taking $g=2\alpha$ leads to the equation  $2\alpha=1+2^{n-3}(k-4)$ which only has $(n,k)=(3,2\alpha+3)$ as positive integer solutions. However, now we can choose $\alpha$ high enough such that $k>k_{\mathrm{max}}$ leading to a contradiction. It is worth noticing that more complicated mechanical systems can be considered to realize every possible oriented surface (see \cite{jordan2001compact} and references therein).

\mbox{}

\section{Example of a non-orientable C-space}\label{appendix more cspaces}
 
\begin{figure}[!ht]\centering
	\includegraphics[width=.85\textwidth]{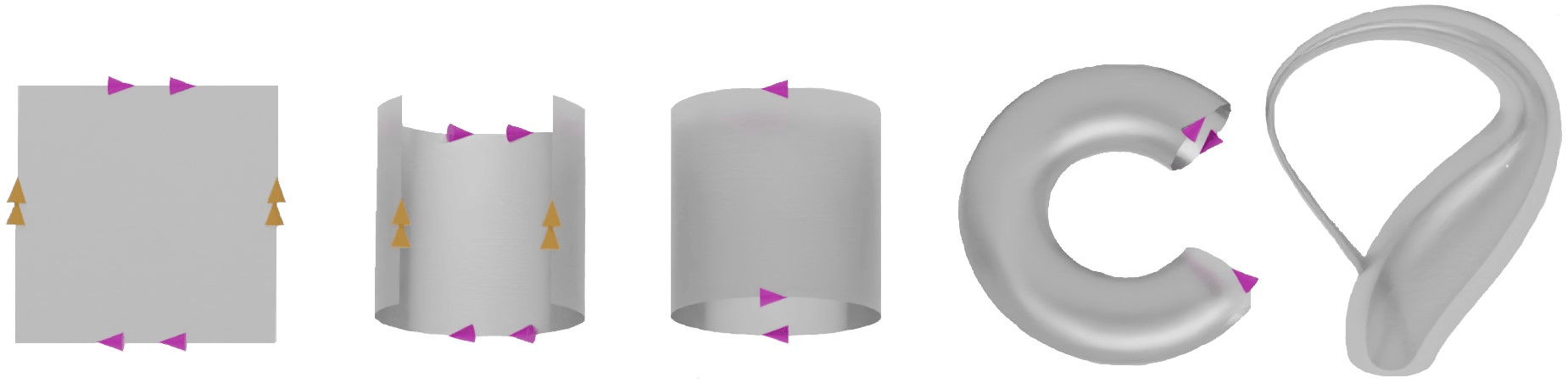}
	\caption{Gluing process to obtain a Klein bottle.}
	\label{fig:GlueKlein}
\end{figure}

By introducing some small changes in the double pendulum, we can change the C-space from a torus (orientable) to a Klein bottle (non-orientable). Recall that a circle $\S^1$ is a closed interval $[0,2\pi]$ where the endpoints are glued. Hence, a torus $\S^1\times\S^1$ can be seen as a square $[0,2\pi]\times[0,2\pi]$ where we first glue the vertical sides $\{0\}\times[0,2\pi]$ and $\{2\pi\}\times[0,2\pi]$ with the same orientation to obtain a cylinder. Afterward, we glue the horizontal ones  $[0,2\pi]\times\{0\}$ and $[0,2\pi]\times\{2\pi\}$ (the boundaries of the cylinder), again with the same orientation, to obtain the torus. This process is shown in Fig.~\ref{fig:GlueTorus}. If we change the orientation of one of the horizontal sides, when we glue the vertical sides, we obtain a cylinder with boundaries with opposite orientations. Gluing them leads to a Klein bottle, as shown in Fig.~\ref{fig:GlueKlein}.

\begin{figure}[!ht]\centering
	\centering
		\includegraphics[width=.75\textwidth]{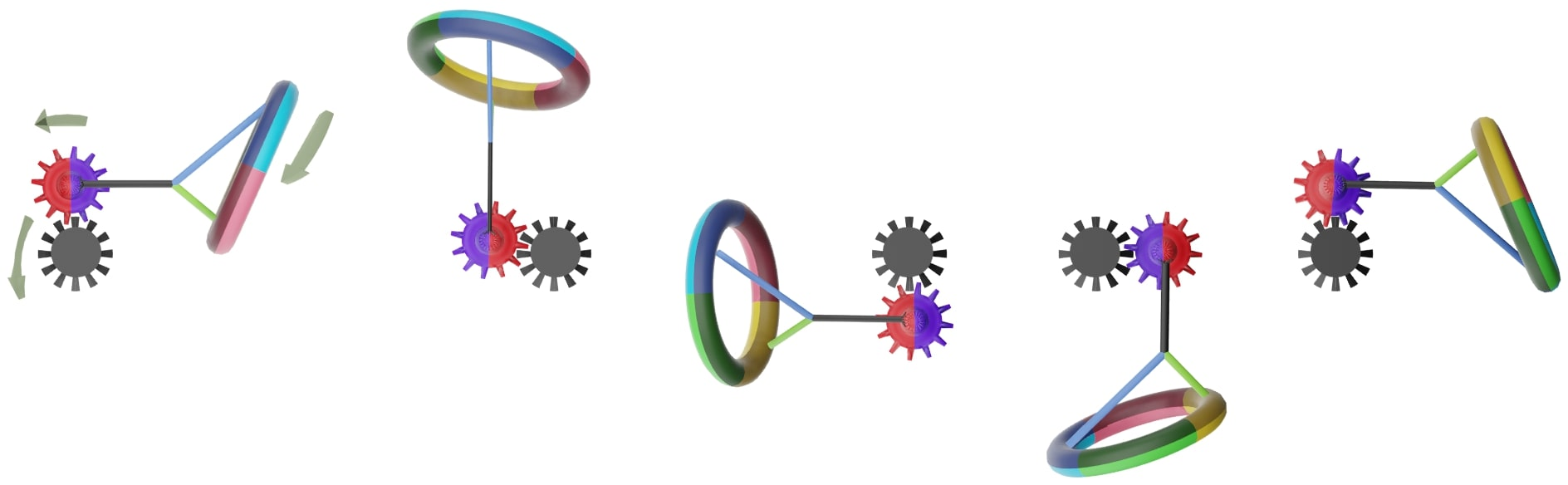}\label{fig. eggbeater}
	\caption{The center gear (black) is fixed, and the other one (purple and red) can turn around the fixed one through the gears. This corresponds to the first pendulum. We then attach a rod to the second gear that can turn freely, corresponding to the second pendulum. The rod can also spin around its axis (shown by the coloured circle at the end), but not freely. The position of the first pendulum governs its spinning: after a full cycle, the circle has turned 180º.}
	\label{fig:eggbeater}
\end{figure}

To realize a Klein bottle in a mechanical system, we add to a double pendulum some gears  so that the second pendulum spins along its own axis when the angle of the \textit{first} pendulum varies (see 
Fig.~\ref{fig:eggbeater}).

If after one whole turn by the first pendulum (without moving the second one), the second one has twist half-turn along its axis, then the vertical sides $\{0\}\times[0,2\pi]$ and $\{2\pi\}\times[0,2\pi]$ will have different orientation leading to a Klein bottle. This construction was actually described, to the best of our knowledge for the first time, by Richard L.W. Brown \cite{brown1973klein}. Brown studies in depth this system, which he refers to as the \textit{egg-beater}, to prove that the C-space is a Klein bottle.

\end{appendices}

\small

\bibliographystyle{style}
\bibliography{ref}

\end{document}